\documentclass[draft]{agujournal2019}
\usepackage{url} 
\usepackage{lineno}
\usepackage{amsmath}
\usepackage[inline]{trackchanges} 
\usepackage{soul}
\draftfalse

\journalname{Earth and Planetary Physics}

\begin{document}

%
%


\title{Hybrid-Vlasov simulation of soft X-ray emissions at the Earth's dayside magnetospheric boundaries}

%
%




\authors{M.~Grandin\affil{1}, H.~K.~Connor\affil{2}, S.~Hoilijoki\affil{1}, M.~Battarbee\affil{1}, Y.~Pfau-Kempf\affil{1}, U.~Ganse\affil{1}, K.~Papadakis\affil{1}, and M.~Palmroth\affil{1,3}}

\affiliation{1}{Department of Physics, University of Helsinki, Helsinki, Finland}
\affiliation{2}{NASA Goddard Space Flight Center, Greenbelt, MD, 20771, USA}
\affiliation{3}{Space and Earth Observation Centre, Finnish Meteorological Institute, Helsinki, Finland}





\correspondingauthor{Maxime Grandin}{maxime.grandin@helsinki.fi}




\begin{keypoints} 
\item We produce soft X-ray images of near-Earth space with a global hybrid-Vlasov simulation with southward interplanetary magnetic field
\item Flux transfer events can produce X-ray signatures despite being transient phenomena if they cumulatively increase the proton density locally
\item Mirror-mode structures in the magnetosheath can also produce soft X-ray signatures in time-integrated images
\end{keypoints}

%
%

%
%


\begin{abstract} 
Solar wind charge exchange produces emissions in the soft X-ray energy range which can enable the study of near-Earth space regions such as the magnetopause, the magnetosheath and the polar cusps by remote sensing techniques. The Solar wind--Magnetosphere--Ionosphere Link Explorer (SMILE) and Lunar Environment heliospheric X-ray Imager (LEXI) missions aim to obtain soft X-ray images of near-Earth space thanks to their Soft X-ray Imager (SXI) instruments. While earlier modeling works have already simulated soft X-ray images as might be obtained by SMILE SXI during its mission, the numerical models used so far are all based on the magnetohydrodynamics description of the space plasma. To investigate the possible signatures of ion-kinetic-scale processes in soft X-ray images, we use for the first time a global hybrid-Vlasov simulation of the geospace from the Vlasiator model. The simulation is driven by fast and tenuous solar wind conditions and purely southward interplanetary magnetic field. We first produce global X-ray images of the dayside near-Earth space by placing a virtual imaging satellite at two different locations, providing meridional and equatorial views. We then analyze regional features present in the images and show that they correspond to signatures in soft X-ray emissions of mirror-mode wave structures in the magnetosheath and flux transfer events (FTEs) at the magnetopause. Our results suggest that, although the time scales associated with the motion of those transient phenomena will likely be significantly smaller than the integration time of the SMILE and LEXI imagers, mirror-mode structures and FTEs can cumulatively produce detectable signatures in the soft X-ray images. For instance, a local increase by 30\% in the proton density at the dayside magnetopause resulting from the transit of multiple FTEs leads to a 12\% enhancement in the line-of-sight- and time-integrated soft X-ray emissivity originating from this region. Likewise, a proton density increase by 14\% in the magnetosheath associated with mirror-mode structures can result in an enhancement in the soft X-ray signal by 4\%. These are likely conservative estimates, given that the solar wind conditions used in the Vlasiator run can be expected to generate weaker soft X-ray emissions than the more common denser solar wind. These results will contribute to the preparatory work for the SMILE and LEXI missions by providing the community with quantitative estimates of the effects of small-scale, transient phenomena occurring on the dayside.\end{abstract}

\section{Introduction}

Over the past two decades, interest has grown in studying near-Earth space by observing the soft X-ray emissions created by charge-exchange interactions between heavy, multiply charged ions and neutral species. For example, charge-exchange interactions between O$^{7+}$ or O$^{8+}$ ions present in the solar wind and neutral hydrogen atoms from the Earth's exosphere are known to lead to photon emissions through the de-excitation of the product ion species, which include photon energies in the soft X-ray range (0.5--0.7~keV). This process is known as solar wind charge exchange (SWCX). A review by \citeA{Sibeck2018} provides in-depth details on the processes at play as well as pioneering works on the study of terrestrial and planetary space through the observation of soft X-ray emissions.

Currently, several satellite missions aim at imaging near-Earth space in soft X-rays. Two major upcoming missions making use of SWCX soft X-ray imaging to reseach near-Earth space are the Lunar Environment heliospheric X-ray Imager (LEXI), led by the Boston University in collaboration with NASA GSFC and several universities (\url{http://sites.bu.edu/lexi}), and the Solar wind--Magnetosphere--Ionosphere Link Explorer (SMILE) mission, designed jointly by the European Space Agency and the Chinese Academy of Sciences \cite{BranduardiRaymont2018}. Both of these missions will provide soft X-ray images of the polar cusps and the magnetosheath, which are known to be the most prominent sources of SWCX soft X-ray emissions in near-Earth space, with a goal of understanding global interaction between the solar wind and the Earth's magnetosphere. After its expected launch in 2024, LEXI will observe the dayside magnetosheath from the lunar surface for up to two weeks of a short mission period due to the harsh lunar environment. SMILE will be launched in 2025 into a highly elliptical polar orbit with an apogee of $\sim$20~Earth radii ($R_\mathrm{E}$), observing the dayside magnetosheath and cusps for up to 40~continuous hours per orbit during three years of its mission period. 

While there have been no wide-field-of-view soft X-ray observations of the geospace, the existence of near-Earth SWCX soft X-ray emissions has been studied in various works using data from the XMM and ROSAT astrophysics missions \cite<e.g.>[]{Carter2010,Carter2011,Cravens2001,Snowden2004,Connor2019,Jung2022,Zhang2022}. Subsequently, modeling efforts were undertaken to understand the soft X-ray emissions detected by the ROSAT mission \cite{Snowden1995} and to quantify the contributions of various source mechanisms \cite<e.g.,>{Cox1998,Cravens2001}. \citeA{Robertson2003_GRL} were the first to produce images of the dayside magnetopause and magnetosheath in soft X-ray emissions through the SWCX mechanism using a numerical model, and based on their results they suggested that it might be possible to make use of those emissions to monitor the solar wind--magnetosphere interactions through remote sensing observations. The same authors also produced simulated images in soft X-rays for a virtual instrument placed on the Earth's surface and observing various directions in the sky \cite{Robertson2003_JGR}, and investigated the polar cusps' signatures in soft X-ray images in further simulations \cite{Robertson2006}.

More recent studies have taken up the modeling efforts as part of the preparatory phases of the LEXI and SMILE missions. \citeA{Collier2018} determined that the surface of the magnetopause could in principle be determined by soft X-ray imaging of near-Earth space, as the line-of-sight-integrated soft X-ray emissions maximize for observations tangential to the magnetopause surface. Based on these results, \citeA{Sun2020} developed a tangent-fitting method to derive the magnetopause position in anticipated images from the Soft X-ray Imager (SXI) instrument onboard SMILE. Alternative methods to determine the magnetopause shape and position from SXI observations have been devised by \citeA{Wang2022}. Among the processes taking place in near-Earth space and which could be studied using soft X-ray imaging, \citeA{Sun2019} showed that SXI will be able to monitor the motions of the dayside magnetopause and of the polar cusps in response to changes in the solar wind driving conditions. \citeA{Connor2021} also addressed similar research questions by simulating soft X-ray images which could be obtained during the SMILE and LEXI missions in an event when the interplanetary magnetic field (IMF) turns southward and an event when the solar wind density abruptly increases. They demonstrated that it might be possible to get information on the time variations in the magnetic reconnection rate at the dayside magnetopause using such remote-sensing techniques, and they argued that these missions could enable the tracing of the energy flow from the solar wind all the way to the cusps with the same instrument.

The above-mentioned modeling studies all used numerical models of space plasma based on the magnetohydrodynamics (MHD) paradigm, which treats the plasma as a fluid. To complement the results obtained with this approach, it would be interesting to carry out modeling studies of SWCX soft X-ray imaging based on a global model of near-Earth space relying on a kinetic description of the plasma. Indeed, a certain number of space plasma processes are intrinsically of kinetic origin, and therefore cannot emerge in MHD simulations. Examples of such processes include the growth of instabilities, wave--particle interactions as well as ion-scale physics associated with magnetic reconnection.

In this paper, we will for the first time carry out a study of soft X-ray emissions relying on a 3D global hybrid-Vlasov simulation. We will make use of the kinetic nature of the model to investigate to what extent near-Earth space processes associated with ion-scale physics might produce signatures in soft X-ray images as will be obtained by the SMILE SXI instrument. To that aim, we will make use of Vlasiator \cite{Palmroth2018}, which is a global hybrid-Vlasov model of near-Earth space. In particular, we will quantify the signatures in terms of line-of-sight- and time-integrated soft X-ray emissions of two transient phenomena occurring on the dayside: mirror-mode waves in the magnetosheath and flux transfer events (FTEs) produced by magnetic reconnection at the dayside magnetopause. Mirror-mode waves are generated by temperature anisotropy in the magnetosheath and appear as nonpropagating structures (in the plasma frame) with anticorrelated variations in plasma density and magnetic field magnitude \cite{Hasegawa1969}. FTEs are flux-rope-like structures forming along the dayside magnetopause in presence of multiple or bursty reconnection lines associated with southward IMF conditions, which often form near the subsolar point and propagate toward the polar cusps \cite{RussellElphic1978,Rijnbeek1984, Berchem1984}.

The paper is organized as follows. The Vlasiator model and the methodology to construct soft X-ray images in the utilized simulation run are introduced in Sect.~\ref{sec:methods}. Then, in Sect.~\ref{sec:images}, we present examples of global images of the dayside near-Earth space obtained with virtual imaging spacecraft placed at two locations: on the dawnside (meridional view) and above the north pole (equatorial view). Section~\ref{sec:transient} describes the analysis and results on the detectability of transient processes (FTEs and mirror-mode waves) in the simulated soft X-ray images. A discussion of the results is provided in Sect.~\ref{sec:discussion}, and Sect.~\ref{sec:conclusion} gives a summary of the main conclusions of the paper.

\section{Models and Methods}
\label{sec:methods}
\subsection{Vlasiator}

\subsubsection{Model}
Vlasiator \cite{vonAlfthan2014,Palmroth2018} is a global hybrid-Vlasov code simulating near-Earth plasma at ion-kinetic scales. In the hybrid-Vlasov approach, ions are described through their velocity distribution function (VDF) which is discretized in a 3-dimensional (3D) velocity grid, resulting in ion populations evolving in a 6D phase space (3D in ordinary space + 3D in velocity space). In practice, Vlasiator solves the Vlasov equation for ions and treats electrons as a massless, charge-neutralizing fluid. While most existing Vlasiator simulation runs consider protons as the sole ion species, heavier ions such as He$^{2+}$ have also been included in a few runs \cite<e.g.,>{Battarbee2020_helium}.

Electromagnetic fields are propagated in time by solving Maxwell's equations under the Darwin approximation, i.e., neglecting the displacement current term in Amp\`ere's law. The equation system is closed through a generalized Ohm's law including the Hall term and a polytropic (adiabatic) description of the electron pressure gradient term. The model considers an ideal geomagnetic dipole field using a nonscaled strength $8\cdot10^{15}\,\mathrm{Tm}^3$ and with a zero tilt between its axis and the $Z$ direction in the Geocentric Solar Ecliptic (GSE) frame of reference --- in the GSE frame, the Earth's center is at the origin, the $X$ axis points toward the Sun, the $Z$ axis toward the north, and the $Y$ axis completes the orthonormal frame and points toward dusk. The geomagnetic dipole is described through a vector potential, which is scaled to zero at the inflow boundary in order to prevent magnetic divergence entering the simulation domain. Magnetic and electric fields are propagated using a finite difference upwind field solver \cite{Londrillo2004}.

The simulation domain extent varies from one run to another, but it generally includes the dayside magnetosphere, the magnetosheath, the bow shock, the ion foreshock (if driving conditions allow for its existence), and part of the magnetotail. This has allowed for a great variety of studies focusing on a wide range of processes, such as foreshock waves \cite{Turc2018}, foreshock cavitons \cite{Tarvus2021}, bow shock nonlocality \cite{Battarbee2020_shock}, magnetosheath waves \cite{Hoilijoki2016,Dubart2020}, energy transfer across the magnetopause \cite{AlaLahti2022}, magnetotail current sheet flapping \cite{Juusola2018}, or auroral proton precipitation \cite{Grandin2019,Grandin2020}. While those past studies were based on 2D--3V runs (2D in ordinary space, 3D in velocity space), recent code developments have enabled the production of the first Vlasiator 3D--3V runs. This study makes use of one such 3D--3V run, described below.

\subsubsection{Simulation Run}
The Vlasiator run used in this study has its simulation domain extending from $-110\,R_\mathrm{E}$ to $50\,R_\mathrm{E}$ in the $X$ direction (with $R_\mathrm{E}$ the Earth's radius; $R_\mathrm{E} = 6371$~km) and confined within $|Y| < 58\,R_\mathrm{E}$ and $|Z| < 58\,R_\mathrm{E}$. The inner boundary lies at $4.7\,R_\mathrm{E}$ from the origin; it is implemented as a perfectly conducting sphere on which VDFs are fixed Maxwellian distributions. External boundaries have Neumann boundary conditions, except for the $+X$ wall from which the solar wind and IMF enter the simulation domain.

Driving conditions in this 3D--3V run are as follows: purely southward IMF with $B_x = B_y = 0$ and $B_z = -5$~nT, solar wind with proton number density of 1~cm$^{-3}$, speed along the $-X$ direction at 750~km$\,$s$^{-1}$, temperature of 500~kK. Solar wind conditions are homogeneous and constant throughout the simulation. Protons are the sole ion species in this run. Initial conditions are such that the whole simulation domain is filled with solar-wind-like Maxwellian VDFs and the superposition of the dipole geomagnetic field with the IMF, and the near-Earth space regions thus form self-consistently during the first few hundred seconds of the simulation. In this study, we will focus on the time interval starting at $t=800$~s, when the dayside magnetosphere is well formed and lasting until the end of the simulation at $t=1506$~s. In this run, an output file was written at a cadence of 1~s, which contains the plasma bulk parameters as well as electromagnetic field components in every simulation cell.

One development which made 3D--3V runs possible was the implementation of  adaptive mesh refinement (AMR) for the ordinary-space mesh. In this run, least-refined regions have a base grid with 8000~km resolution, and there are three refinement levels at 4000, 2000 and 1000~km resolution to improve the description of regions of interest where ion-scale kinetic processes are important (bow shock, magnetosheath, magnetopause, magnetotail current sheet). At such resolutions, one expects that some of the kinetic processes cannot arise, as the ion inertial length is not resolved. For instance, the ion cyclotron instability cannot grow with such a coarse grid, hence most of the free energy associated with unstable ion distributions will feed the mirror instability \cite{Dubart2020}. Nevertheless, as will be seen in the upcoming sections, the general morphology of the near-Earth space regions and the plasma parameter values are realistic, making this run usable for the purpose of this study. The velocity space is a uniform 3D Cartesian grid with a resolution of 40~km$\,$s$^{-1}$. In the solar wind, the Alfvén speed is 109~km\,s$^{-1}$ and the thermal speed is 111~km\,s$^{-1}$, and in regions of near-Earth space which are hotter or have a stronger magnetic field, these characteristic speeds are larger, ensuring that the associated physical processes are resolved in a satisfactory way in the regions of interest. The electric and magnetic fields are solved on a uniform Cartesian grid at constant resolution of 1000~km in a process described by \citeA{Papadakis2022}.

\begin{figure}
   \noindent\includegraphics[width=\textwidth]{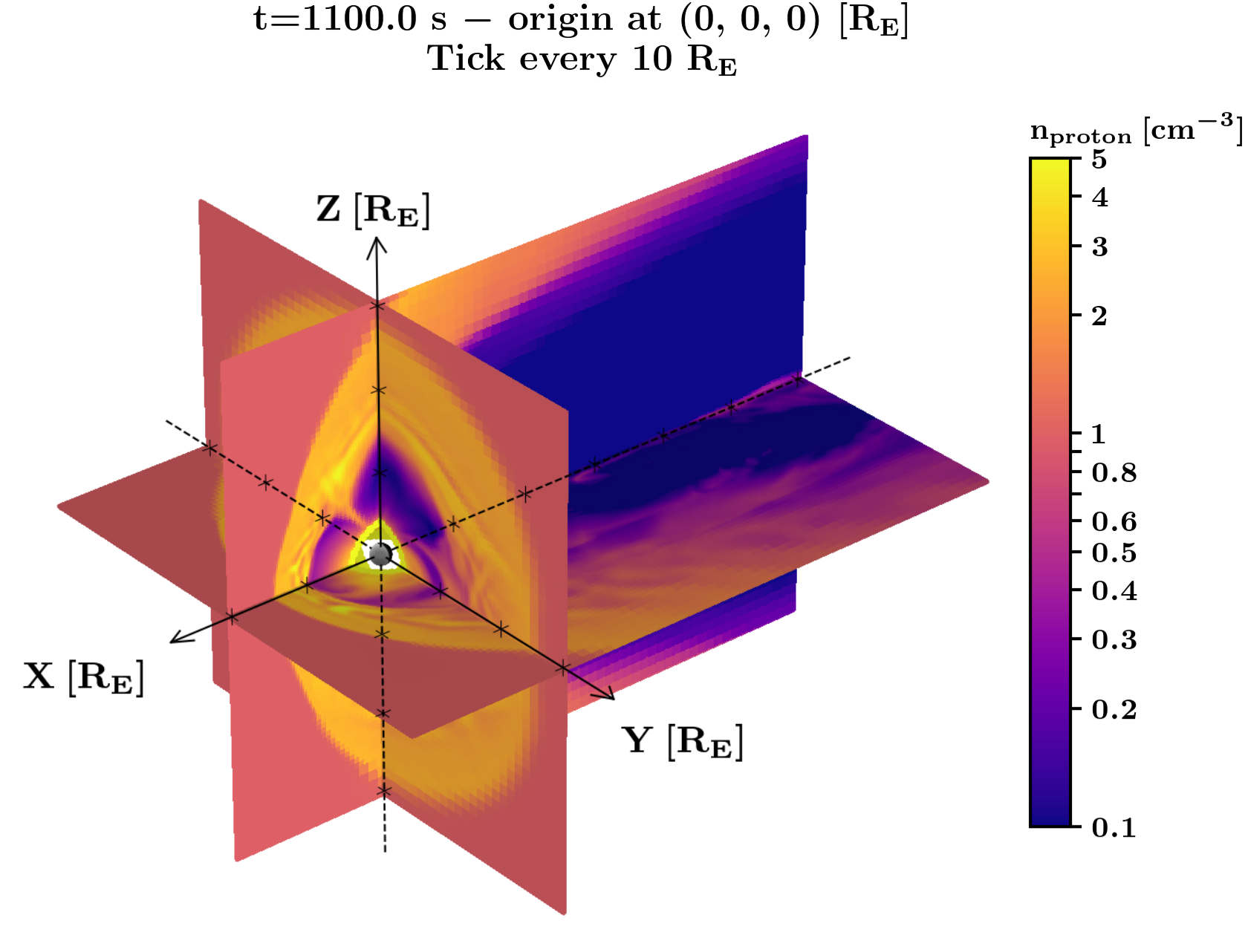}
   \caption{Proton number density in the $X=0$, $Y=0$ and $Z=0$ planes of the simulation domain at $t = 1100$~s in the Vlasiator simulation, wherein the bow shock, the magnetopause and the northern polar cusp are prominent. One can also identify wave-like structures in the dayside magnetosheath.}
   \label{fig:run_overview}
\end{figure}

Figure~\ref{fig:run_overview} gives an overview of the simulation domain at $t=1100$~s. It consists of three slices in the $X=0$, $Y=0$ and $Z=0$ planes showing the proton number density. While the chosen viewing angle and slice extents enable seeing dayside structures and processes relevant to this study, the actual simulation domain extends beyond this figure (see above). In the figure, one can identify the bow shock, the magnetosheath exhibiting wave-like structures, the magnetopause as well as the northern polar cusp. The magnetotail, partly hidden behind the $X=0$ slice, will not be the focus of this study but exhibits complex dynamics.

\subsection{Soft X-ray Image Generation}
\subsubsection{Analytical Expression of Soft X-ray Emissivity}

In this study, the method to derive the local soft X-ray emissivity within the simulation will follow the same approach as past studies \cite<e.g.>{Connor2021,Sun2020}. It is calculated at position $\mathbf{r}$ in the simulation domain with the following expression
\begin{linenomath*}
  \begin{equation}\label{eq:Qloc}
    Q_\mathrm{loc}(\mathbf{r}) = \dfrac{\alpha_X}{4 \pi} n_\mathrm{p}(\mathbf{r}) n_\mathrm{H}(\mathbf{r}) V_\mathrm{eff}(\mathbf{r}),
  \end{equation}
\end{linenomath*}
where $\alpha_X$ is the interaction efficiency factor, $n_\mathrm{p}$ is the proton number density, $n_\mathrm{H}$ is the neutral hydrogen atom density, and $V_\mathrm{eff}$ is the so-called ``effective velocity'', defined as
\begin{linenomath*}
  \begin{equation}\label{eq:Veff}
    V_\mathrm{eff}(\mathbf{r}) = \sqrt{V_\mathrm{p}(\mathbf{r})^2 + \dfrac{5}{3}\dfrac{k_B T(\mathbf{r})}{m_\mathrm{p}}}.
  \end{equation}
\end{linenomath*}
In Eq.~\eqref{eq:Veff}, $V_\mathrm{p}$ is the proton bulk velocity, $k_B$ is Boltzmann's constant, $T$ is the plasma temperature, and $m_\mathrm{p}$ is the proton mass. This equation expresses $V_\mathrm{eff}$ in m$\,$s$^{-1}$. Note that protons do not emit soft X-rays.  The highly charged, heavy solar wind ions like C$^{6+}$, N$^{6+}$, N$^{7+}$, Ne$^{9+}$, S$^{10+}$, O$^{7+}$, and O$^{8+}$ emit soft X-rays through SWCX \cite{Sibeck2018}. By multiplying the interaction efficiency factor $\alpha_X$, the proton-based quantity in Eq.~\eqref{eq:Qloc} is transformed into the soft X-ray emissivity caused by the source ions. 

Since Vlasiator does not simulate the exosphere, we use the analytical model from \citeA{Cravens2001} for the neutral density, given as
\begin{linenomath*}
  \begin{equation}\label{eq:nH}
    n_\mathrm{H}(\mathbf{r})= 25 \left( \dfrac{10\,R_\mathrm{E}}{r} \right)^3,
  \end{equation}
\end{linenomath*}
with $r$ the distance of the considered location to the Earth's center. The above expression gives $n_\mathrm{H}$ in cm$^{-3}$, and in the continuation of the study, we will express soft X-ray emissivity quantities using centimeter as the unit of length, following the common usage for this specific application. For instance, $Q_\mathrm{loc}$ will be expressed in keV\,cm$^{-3}\,$s$^{-1}$\,sr$^{-1}$. In this study, we will use an interaction efficiency factor value of $\alpha_X = 1 \times 10^{-15}$~eV\,cm$^2$, following \citeA{Sun2019} and references therein.

\subsubsection{Line-of-Sight Integration from a Virtual Spacecraft}
In order to simulate soft X-ray images close to as they would be obtained by a spacecraft such as LEXI and SMILE, we place a virtual spacecraft in the Vlasiator simulation domain and calculate the line-of-sight-integrated value of the local soft X-ray emissivity along multiple viewing directions within the instrument field-of-view, which corresponds to many pixels in the images. We define this quantity $Q_\mathrm{int}(\varphi,\lambda)$ as a function of the azimuth $\varphi$ and elevation $\lambda$ of a given line of sight.
\begin{linenomath*}
  \begin{equation}
    Q_\mathrm{int}(\varphi,\lambda) = \int Q_\mathrm{loc}(l_{\varphi,\lambda})\,\mathrm{d}l_{\varphi,\lambda},
  \end{equation}
\end{linenomath*}
with $l_{\varphi,\lambda}$ the distance from the spacecraft along the line of sight associated with the \mbox{($\varphi$, $\lambda$)} azimuth--elevation pair. By convention, we will have the \mbox{(0, 0)} pair corresponding to the direction toward the Earth's center. The obtained instantaneous line-of-sight emissivity $Q_\mathrm{int}$ will be given in keV\,cm$^{-2}\,$s$^{-1}$\,sr$^{-1}$ in this study.

Soft X-ray images simulated using this Vlasiator run will have an angular resolution of 0.33$^\circ$ in both azimuth and elevation, and the line-of-sight integration will take place between the virtual satellite location and the outer boundary of the simulation domain. If a given line of sight intersects the inner boundary, $Q_\mathrm{int}$ will not be calculated. The virtual spacecraft will be placed either at \mbox{(0, $-30\,R_\mathrm{E}$, 0)} in GSE coordinates (i.e., observing from the dawnside) or at \mbox{(0, 0, $30\,R_\mathrm{E}$)} (i.e., observing from above the north pole). The two virtual satellites are selected to provide the side and polar views of the Earth's magnetosphere similar to LEXI and SMILE, respectively, while keeping the radial distances fixed at 30~$R_\mathrm{E}$ between the apogees of LEXI and SMILE for easy comparison of soft X-ray signatures in polar and side views. Both instantaneous values of $Q_\mathrm{int}$ and time-integrated ones over 300~s, $Q_\mathrm{int\_300s}$, will be shown, as the latter correspond to the integration time that SMILE and LEXI soft X-ray imagers require for good signal-to-noise ratios \cite{BranduardiRaymont2018,Connor2021}.

\section{Simulated Soft X-ray Images}
\label{sec:images}
In this section, we present the simulated soft X-ray images obtained from the Vlasiator run when the virtual imaging spacecraft is viewing the dayside magnetosphere either from the dawnside or from above the north pole. We first show instantaneous images alongside relevant plasma parameters in the noon--midnight meridional plane or equatorial plane, respectively. Animations showing such instantanous images every second from $t = 800$~s to $t=1506$~s are provided in supplementary material. We then time-integrate these instantaneous images during three 300~s time intervals and discuss the main features that can be observed.

\subsection{Instantaneous Simulated Soft X-ray Images}
\subsubsection{View from Dawn}

\begin{figure}
   \noindent\includegraphics[width=0.9\textwidth]{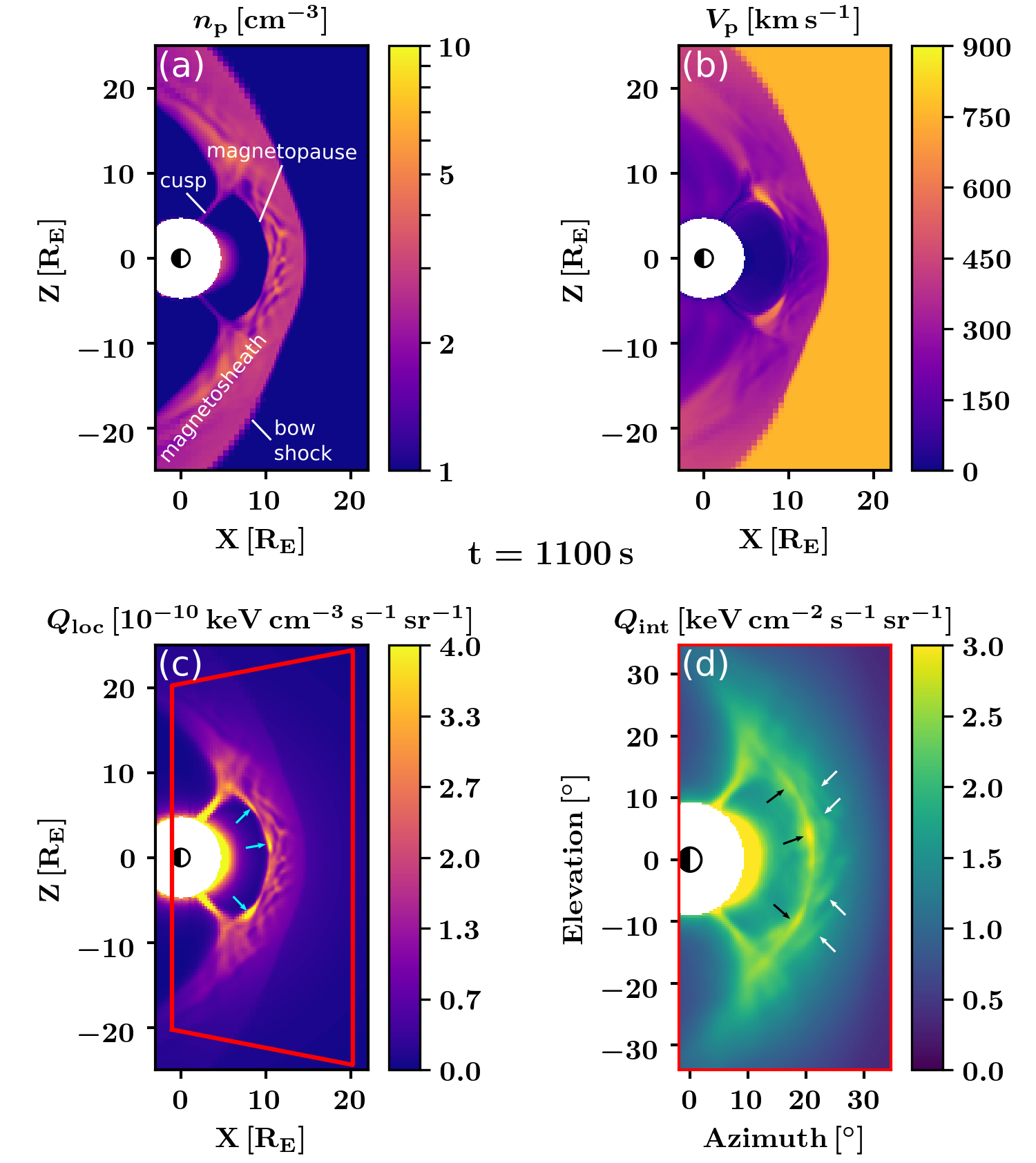}
   \caption{(a)~Proton number density, (b)~proton bulk velocity, and (c) local soft X-ray emissivity in the \mbox{$Y=0$} plane at \mbox{$t=1100$~s}. (d)~Instantaneous soft X-ray image with 1~s integration time from a virtual spacecraft placed at (0, $-30\,R_\mathrm{E}$, 0) at \mbox{$t=1100$~s}. The cyan arrows in panel~(c) indicate flux transfer event signatures in soft X-rays, and the red trapezoid indicates the intersection of the instrument's field of view shown in panel~(d) with the \mbox{$Y=0$} plane. The black and white arrows in panel~(d) indicate structures discussed in the text.}
   \label{fig:snapshot_XZ}
\end{figure}

Figure~\ref{fig:snapshot_XZ} shows plasma parameters in the noon--midnight meridional plane \mbox{($Y=0$)} at $t=1100$~s in the simulation. Figure~\ref{fig:snapshot_XZ}a displays the proton number density, in which the magnetosheath and the polar cusps stand out. One can identify the bow shock, whose subsolar point lies approximately at 15\,$R_\mathrm{E}$. Given the purely southward IMF orientation, the bow shock is essentially quasi-perpendicular, which is why no ion foreshock is present. Within the magnetosheath, density irregularities are particularly prominent. We will show in Sect.~\ref{sec:mirrormode} that these correspond to mirror-mode waves. The magnetopause subsolar point lies at about 10\,$R_\mathrm{E}$.

Figure~\ref{fig:snapshot_XZ}b gives the proton bulk velocity. When crossing the bow shock, the solar wind slows down from its inflow speed of 750~km\,s$^{-1}$ to velocities on the order of 200--300~km\,s$^{-1}$. We can see that plasma convection at the dayside magnetopause gets faster at higher latitudes, before slowing down again when the plasma reaches the polar cusps.

Figure~\ref{fig:snapshot_XZ}c presents the local soft X-ray emissivity in the plane, $Q_\mathrm{loc}$ (calculated with Eq.~\ref{eq:Qloc}), which depends not only on the previous two parameters ($n_\mathrm{p}$ and $V_\mathrm{p}$), but also on the plasma temperature and the neutral density, not shown in the figure. The most prominent features in $Q_\mathrm{loc}$ are the polar cusps and the magnetosheath, with values reaching $4 \times 10^{-10}$~keV\,cm$^{-3}$\,s$^{-1}$\,sr$^{-1}$. In the magnetosheath, the wave field is particularly visible. Brighter areas along the dayside magnetopause can also be identified (indicated with cyan arrows); we will show in Sect.~\ref{sec:FTE} that these are flux transfer events (FTEs), which appear as regions of slightly enhanced proton density in Fig.~\ref{fig:snapshot_XZ}a.

Figure~\ref{fig:snapshot_XZ}d shows the instantaneous image of line-of-sight-integrated soft X-ray emissions, $Q_\mathrm{int}$, viewed from a virtual imaging satellite placed at (0, $-30\,R_\mathrm{E}$, 0), at this same time. Due to the geometry associated with the setup, the field of view associated with this image intersects the noon--midnight meridional plane within the red trapezoid in Fig.~\ref{fig:snapshot_XZ}c. Elevation is the angle along the $Z$ direction, whereas azimuth is along $X$. In this instantaneous image, the brightest areas correspond to the cusps and the magnetopause, with values of $Q_\mathrm{int}$ on the order of 3~keV\,cm$^{-2}$\,s$^{-1}$\,sr$^{-1}$. The cusps are particularly bright because of they have a high density of neutrals (see Eq~\eqref{eq:nH}). It was shown in previous studies that lines-of-sight with maximum brightness correspond to directions tangential to the magnetopause \cite{Collier2018}. The very bright area close to the inner boundary is caused by boundary effects, as proton density is elevated at low latitudes near the inner boundary (Fig.~\ref{fig:snapshot_XZ}a), due to leakage of cold plasma resulting from the boundary condition. However, the bright SWCX soft X-ray emission near the inner boundary does not exist in reality and is an artificial effect by calculating soft X-ray emissivity based on proton parameters (Eq.~1).

In the instantaneous image of $Q_\mathrm{int}$, one can notice signatures likely associated with the wave field in the magnetosheath (indicated with white arrows), as well as brighter spots along the dayside magnetopause (indicated with black arrows). This means that, despite the line-of-sight integration and parallax effects, some of the structures identified in $Q_\mathrm{loc}$ (Fig.~\ref{fig:snapshot_XZ}c) also show in $Q_\mathrm{int}$. The possible relation between such structures and transient processes will be investigated in Sect.~\ref{sec:transient}.

An animated version of Fig.~\ref{fig:snapshot_XZ} is provided in the supplementary material as Movie~S1. In this animation, we can especially visualize how the magnetosheath wave signatures and dayside magnetopause bright spots move as a function of time. The fact that the latter are colocated with confined regions of increased proton density forming near the subsolar point and transiting along the magnetopause toward the polar cusps strongly suggests that these are signatures of FTEs (see Sect.~\ref{sec:FTE} and discussion in Sect.~\ref{sec:discussion}).

\subsubsection{View from North}

\begin{figure}
   \noindent\includegraphics[width=0.9\textwidth]{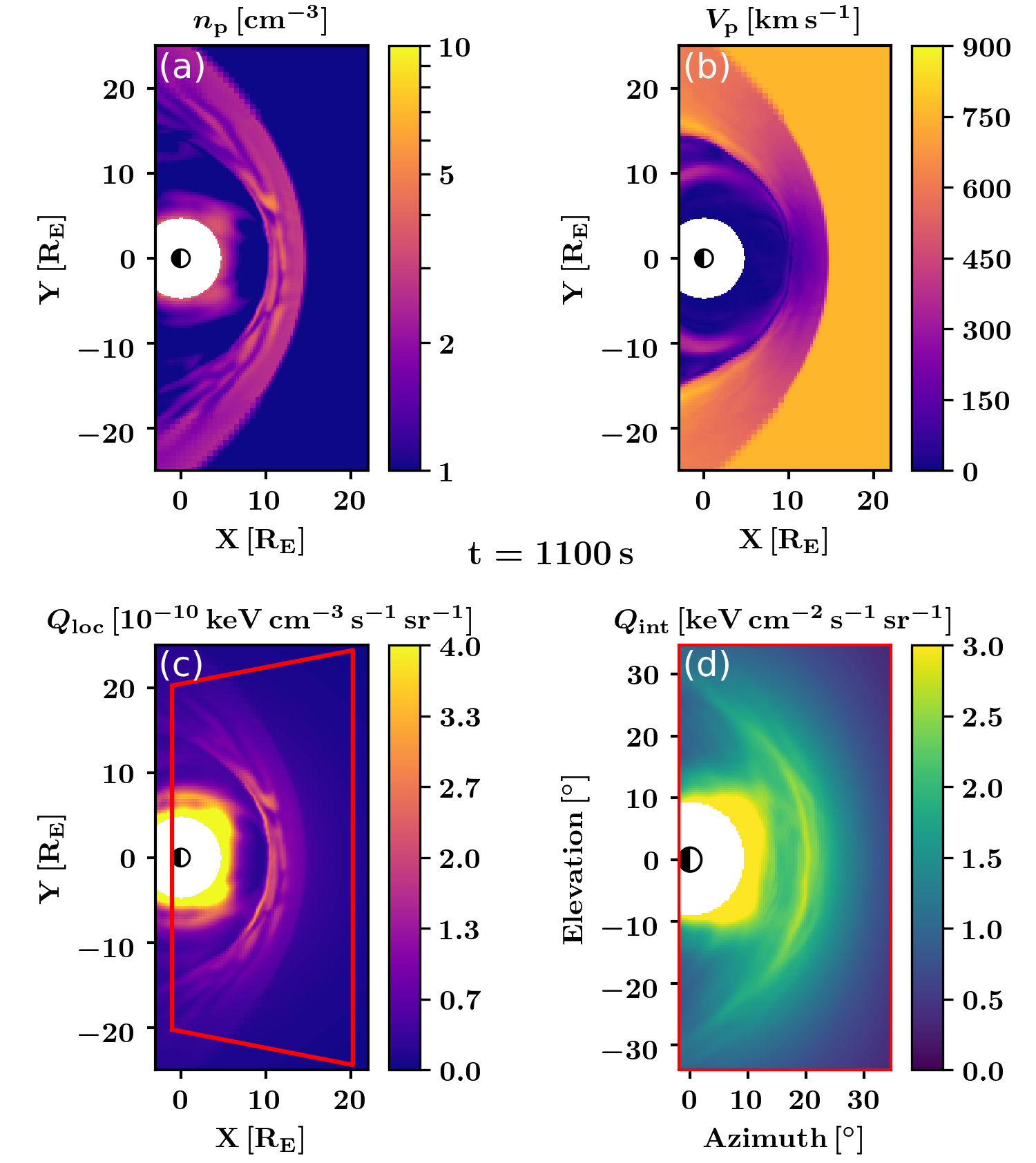}
   \caption{(a)~Proton number density, (b)~proton bulk velocity, and (c) local soft X-ray emissivity in the \mbox{$Z=0$} plane at \mbox{$t=1100$~s}. (d)~Instantaneous soft X-ray image with 1~s integration time from a virtual spacecraft placed at (0, 0, $30\,R_\mathrm{E}$) at \mbox{$t=1100$~s}. The red trapezoid in panel (c) indicates the intersection of the instrument's field of view shown in panel (d) with the \mbox{$Z=0$} plane.}
   \label{fig:snapshot_XY}
\end{figure}

Figure~\ref{fig:snapshot_XY} is analogous to Fig.~\ref{fig:snapshot_XZ}, but this time the virtual imaging spacecraft has been placed above the north pole, at (0, 0, $30\,R_\mathrm{E}$) in GSE coordinates. The chosen time step for this instantaneous snapshot of plasma parameters and soft X-ray emissions is $t=1100$~s, like previously. Figures~\ref{fig:snapshot_XY}a--c show parameters in the equatorial plane \mbox{(Z=0)}, and Fig.~\ref{fig:snapshot_XY}d shows the line-of-sight integrated soft X-ray emissions as a function of azimuth (along the $X$ direction) and elevation (this time along $Y$). 

The bow shock and dayside magnetopause are again prominent in the proton density panel (Fig.~\ref{fig:snapshot_XY}a), as are the magnetosheath waves. The proton bulk velocity panel indicates the abrupt decrease in velocity across the bow shock, while the magnetosheath plasma accelerates once it reaches the flanks (Fig.~\ref{fig:snapshot_XY}b). In local soft X-ray emissivity (Fig.~\ref{fig:snapshot_XY}c), it is again the magnetosheath that shows up the most (ignoring the artifacts near the inner boundary).

In the instantaneous soft X-ray emission image (Fig.~\ref{fig:snapshot_XY}d), a bright arc is visible across elevations, corresponding to the dayside magnetopause observed tangentially. Beyond the magnetopause signature, one can also see elongated structures originating from the magnetosheath. Due to the viewing angle, polar cusps are not visible in this image. Like in Fig.~\ref{fig:snapshot_XZ}, the bright emissions near the inner boundary are associated with boundary effects and would not be present in nature. An animated version of Fig.~\ref{fig:snapshot_XY} is provided in the supplementary material as Movie~S2. In this animation, we can see how the elongated magnetosheath structures drift Earthward until they merge with the magnetopause signature, which occasionally brightens and exhibits undulations (e.g., around $t = 1000$~s and $t=1440$~s).

\subsection{Time-Integrated Soft X-ray Images}
\subsubsection{View from Dawn}

\begin{figure}
   \noindent\includegraphics[width=\textwidth]{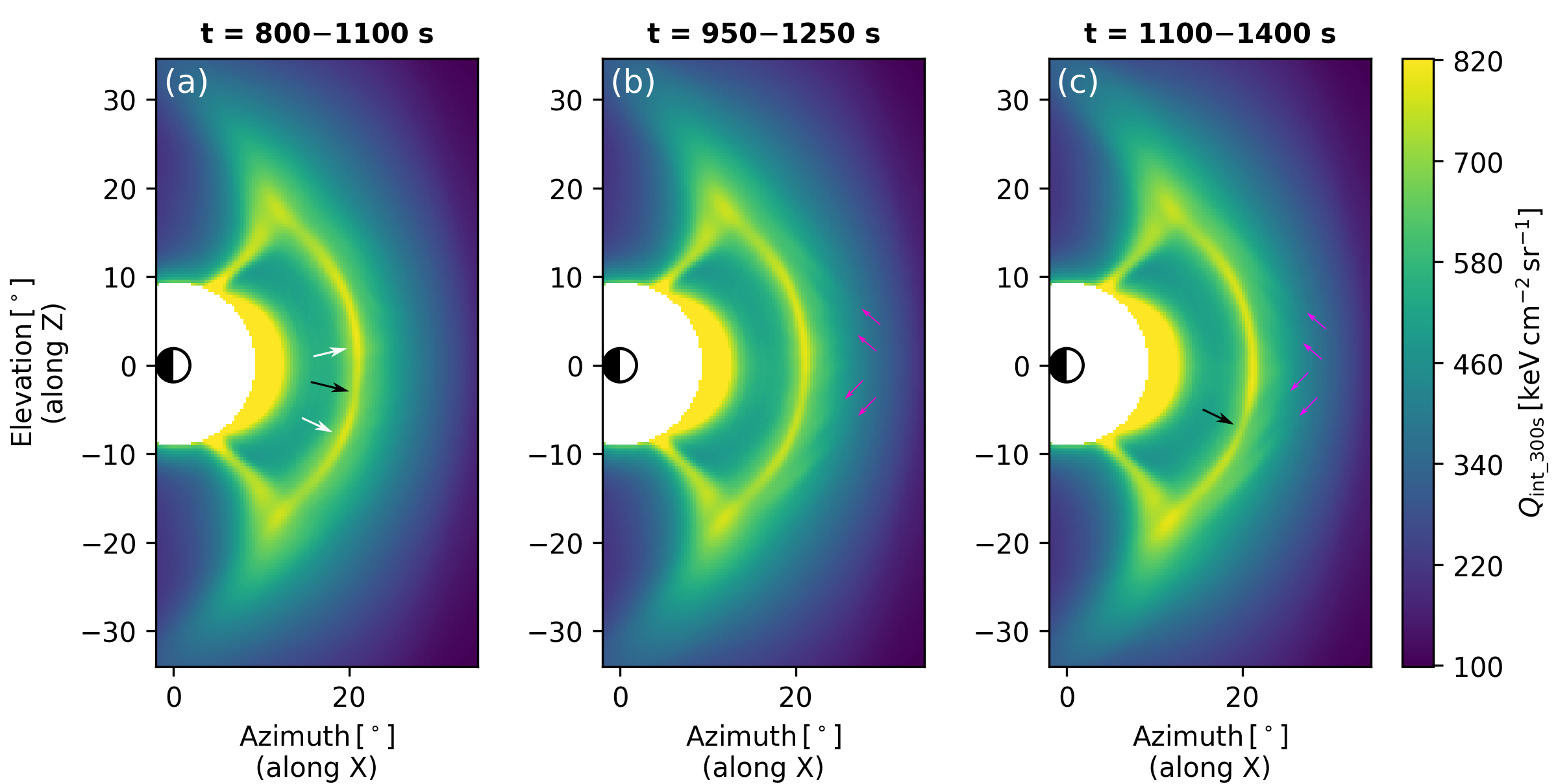}
   \caption{Soft X-ray image obtained from a virtual spacecraft located at $(0, -30\,R_\mathrm{E}, 0)$ with 300~s integration time starting at \mbox{(a)~$t = 800$~s,} \mbox{(b)~$t = 950$~s,} and \mbox{(c)~$t = 1100$~s} in the simulation.}
   \label{fig:Qint300sXZ}
\end{figure}

While the instantaneous soft X-ray images presented above show many interesting features, they do not correspond to images which could be obtained by SMILE and LEXI, as the instruments will need an integration time on the order of 300~s to obtain a sufficient number of counts. We therefore produce time-integrated images over three 300~s time intervals during the studied part of the simulation. We first consider the view from a virtual imaging spacecraft in the dawn sector; Figure~\ref{fig:Qint300sXZ} shows time-integrated soft X-ray images obtained during $t = 800$--1100~s (Fig.~\ref{fig:Qint300sXZ}a), $t = 950$--1250~s (Fig.~\ref{fig:Qint300sXZ}b), and $t = 1100$--1400~s (Fig.~\ref{fig:Qint300sXZ}c). In these images, the brightest structures correspond to the cusps as well as the dayside magnetopause.

It is clear from the figures that the time integration blurs out most small-scale structures that are associated with transient phenomena at the magnetopause and in the magnetosheath. However, one can see some irregularity in the shape and brightness of the magnetopause signature, such as a dimmer and thinner region at elevations near $-3^\circ$ in Fig.~\ref{fig:Qint300sXZ}a, and also at elevations near $-7^\circ$ in Fig.~\ref{fig:Qint300sXZ}c, for instance (indicated with black arrows). Conversely, parts of the dayside magnetopause signature appear brighter, like regions near $+3^\circ$ and $-9^\circ$ elevations in Fig.~\ref{fig:Qint300sXZ}a (indicated with white arrows). We will investigate the possible connection between these brighter $Q_\mathrm{int\_300s}$ values at the dayside magnetopause and FTEs in Sect.~\ref{sec:FTE}.

Besides, Figs~\ref{fig:Qint300sXZ}b--c exhibit brighter stripes in their magnetosheath signatures (indicated with magenta arrows). Such stripes have oblique orientations with respect to the Earth--Sun line and can be seen in both the northern and the southern part of the domain. In Sect.~\ref{sec:mirrormode}, we will show that these signatures are likely the result of mirror-mode waves in the magnetosheath.

\begin{figure}
   \noindent\includegraphics[width=\textwidth]{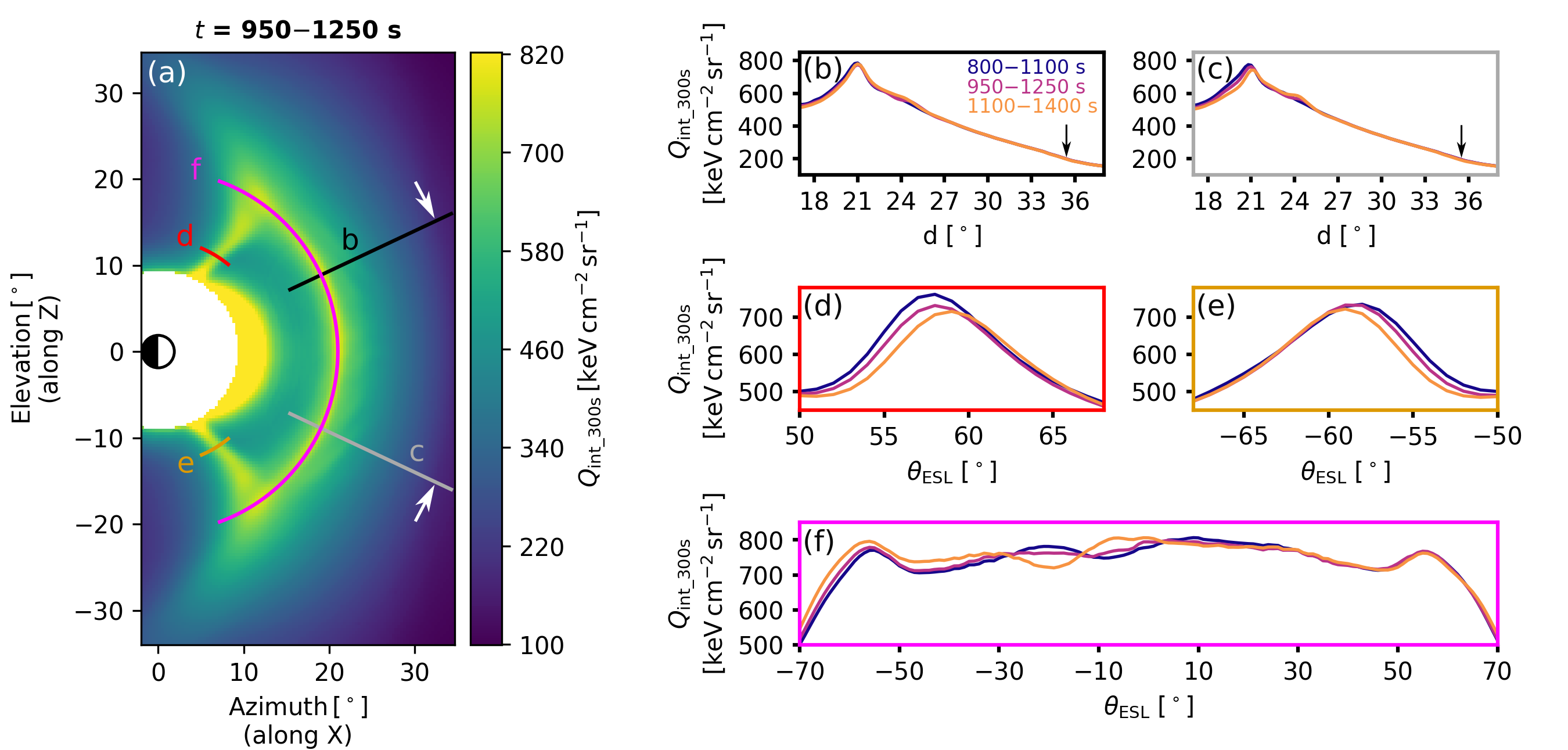}
   \caption{(a)~Time-integrated soft X-ray image from $t=950$ to $t=1250$~s obtained from a virtual spacecraft located at $(0, -30\,R_\mathrm{E}, 0)$ (same as Fig.~\ref{fig:Qint300sXZ}b), with several cuts along and across boundary region signatures. (b)~Soft X-ray emission along the black cut through the northern-hemisphere magnetopause and magnetosheath signatures. (c)~Same along the grey cut through the southern-hemisphere magnetopause and magnetosheath signatures. (d)~Same along the red cut across the northern cusp signature. (e)~Same along the orange cut across the southern cusp signature. (f)~Same along the magenta cut following the dayside magnetopause signature. In panels b--f, the three lines correspond to the three studied 300~s time intervals for image acquisition.}
   \label{fig:cutsXZ}
\end{figure}

In order to better visualize differences between the three images, we will look at $Q_\mathrm{int\_300s}$ values along a few selected cuts. Fig.~\ref{fig:cutsXZ}a reproduces the time-integrated image within $t=950$--1250~s (i.e., Fig.~\ref{fig:Qint300sXZ}b) and indicates where five cuts have been performed, along which the $Q_\mathrm{int\_300s}$ values are extracted and displayed in Figs.~\ref{fig:cutsXZ}b--f for the three time intervals.

Figs.~\ref{fig:cutsXZ}b and \ref{fig:cutsXZ}c show time-integrated soft X-ray emission values along the black and grey straight lines, respectively, which both cut across the magnetopause and through the magnetosheath. The horizontal axis is the angular distance, noted $d$, of a point on the line from the (0, 0) viewing direction (calculated as $\sqrt{\varphi^2+\lambda^2}$). In both panels, $Q_\mathrm{int\_300s}$ peaks at $d \approx 21^\circ$, which corresponds to the tangent direction to the magnetopause. While the curves corresponding to the three integrating time intervals are almost perfectly superimposed on top of each other in Fig.~\ref{fig:cutsXZ}b, one can see that the peak was slightly reduced and drifted to larger $d$ values with time in Fig.~\ref{fig:cutsXZ}c. Around $d \approx 24^\circ$, there are small-amplitude fluctuations visible in the violet and orange curves, in both panels, corresponding to the stripes identified in Figs.~\ref{fig:Qint300sXZ}b--c. A last observation that can be made from these two panels is a change in the slope of $Q_\mathrm{int\_300s}$ occurring at $d \approx 35.5^\circ$, indicated with black arrows. The slope changes from about $-27$ to $-17$~keV\,cm$^{-2}$\,sr$^{-1}$\,deg$^{-1}$. This transition takes place when the line of sight becomes tangent to the (quasi-perpendicular) bow shock, as shown in Fig.~4 of \citeA{Connor2021}. The corresponding locations in Fig.~\ref{fig:cutsXZ}a are indicated with white arrows.

Values of $Q_\mathrm{int\_300s}$ along the red and orange curved lines, cutting through the northern and southern cusp, respectively, are shown in Figs.~\ref{fig:cutsXZ}d and \ref{fig:cutsXZ}e. The horizontal axis is the angle between the Earth--point direction and the Earth--Sun line, noted $\theta_\mathrm{ESL}$. In both hemispheres, the cusp signature drifted slightly poleward with time, although the differences are relatively small.

Figure~\ref{fig:cutsXZ}f give the $Q_\mathrm{int\_300s}$ values along the magenta curve, which essentially follows the magnetopause signature in the images, as a function of $\theta_\mathrm{ESL}$. It shows quite clearly the variations in $Q_\mathrm{int\_300s}$ previously identified in Fig.~\ref{fig:Qint300sXZ} and enables estimating the amplitude of those variations, which can reach up to 90~keV\,cm$^{-2}$\,sr$^{-1}$ (orange curve peak-to-trough amplitude at $\theta_\mathrm{ESL}$ within [$-20^\circ$, $-10^\circ$]), i.e. representing about 10\% of the average value. Other peaks can be identified at $\theta_\mathrm{ESL} \approx \pm55^\circ$, corresponding to locations in the high-altitude polar cusps.

\subsubsection{View from North}

\begin{figure}
   \noindent\includegraphics[width=\textwidth]{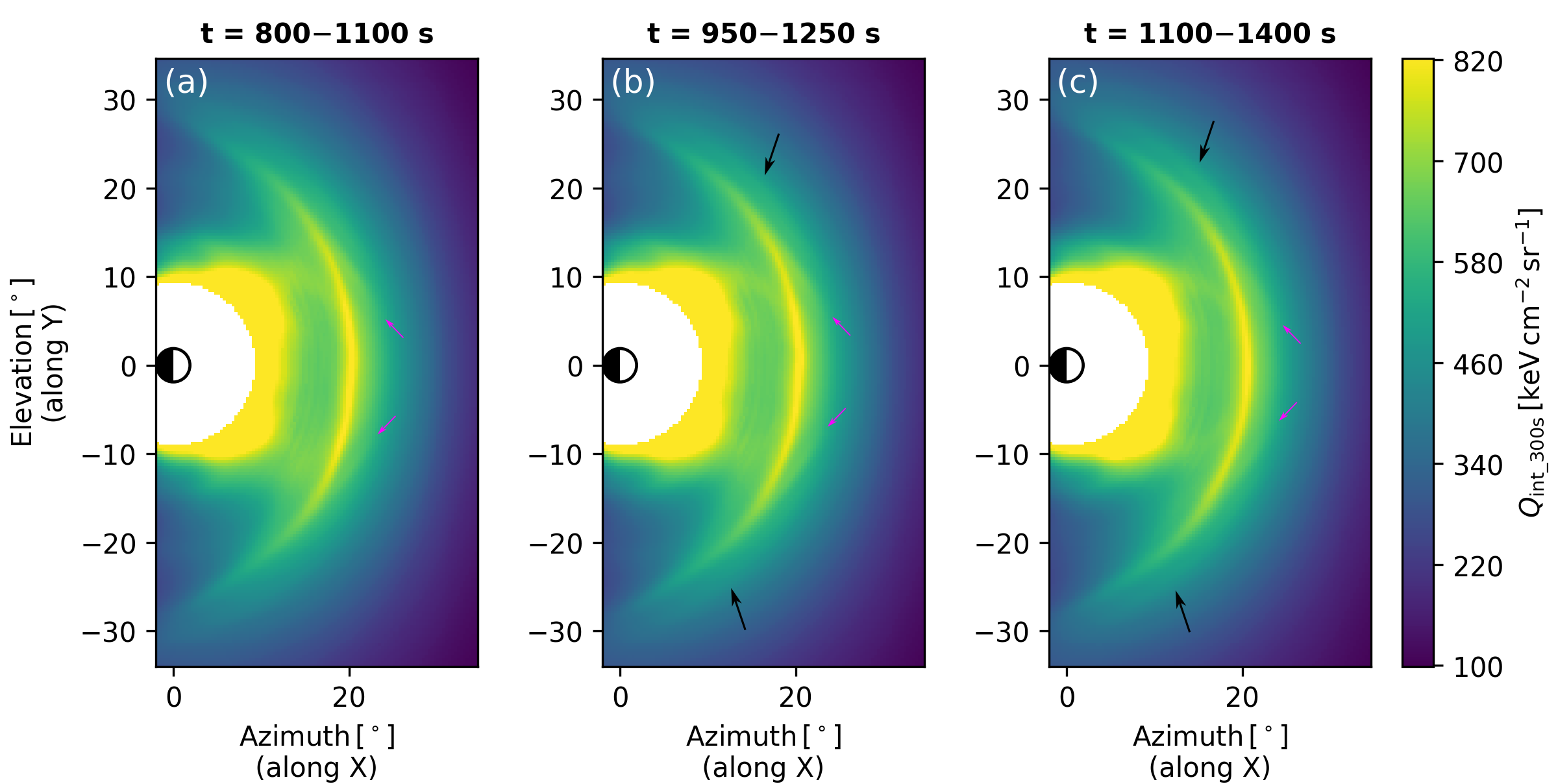}
   \caption{Same as Fig.~\ref{fig:Qint300sXZ} but with the virtual spacecraft placed at $(0, 0, 30\,R_\mathrm{E})$.}
   \label{fig:Qint300sXY}
\end{figure}

By applying the same methodology as above, we can also produce time-integrated soft X-ray images as would be obtained from the virtual imaging spacecraft placed above the north pole. Figure~\ref{fig:Qint300sXY} presents the three images corresponding to the same integration intervals (i.e, $t = 800$--1100~s, $t = 950$--1250~s, and $t = 1100$--1400~s).

In the three panels, the brightest signature comes from the dayside magnetopause, with brightness maximizing at locations corresponding to low nose angles and signatures becoming fainter toward the flanks of the magnetopause. Magnetosheath stripes also show from this perspective, and are visible in all three panels (indicated with magenta arrows), with mostly oblique orientations. Beyond the magnetopause signature near the flanks, a second, fainter line parallel to it can be seen, especially in Figs.~\ref{fig:Qint300sXY}b--c (indicated with black arrows). This parallel line might be associated with the same processes as the stripes closer to the subsolar point, as they appear similar in terms of brightness and orientation.

\begin{figure}
   \noindent\includegraphics[width=\textwidth]{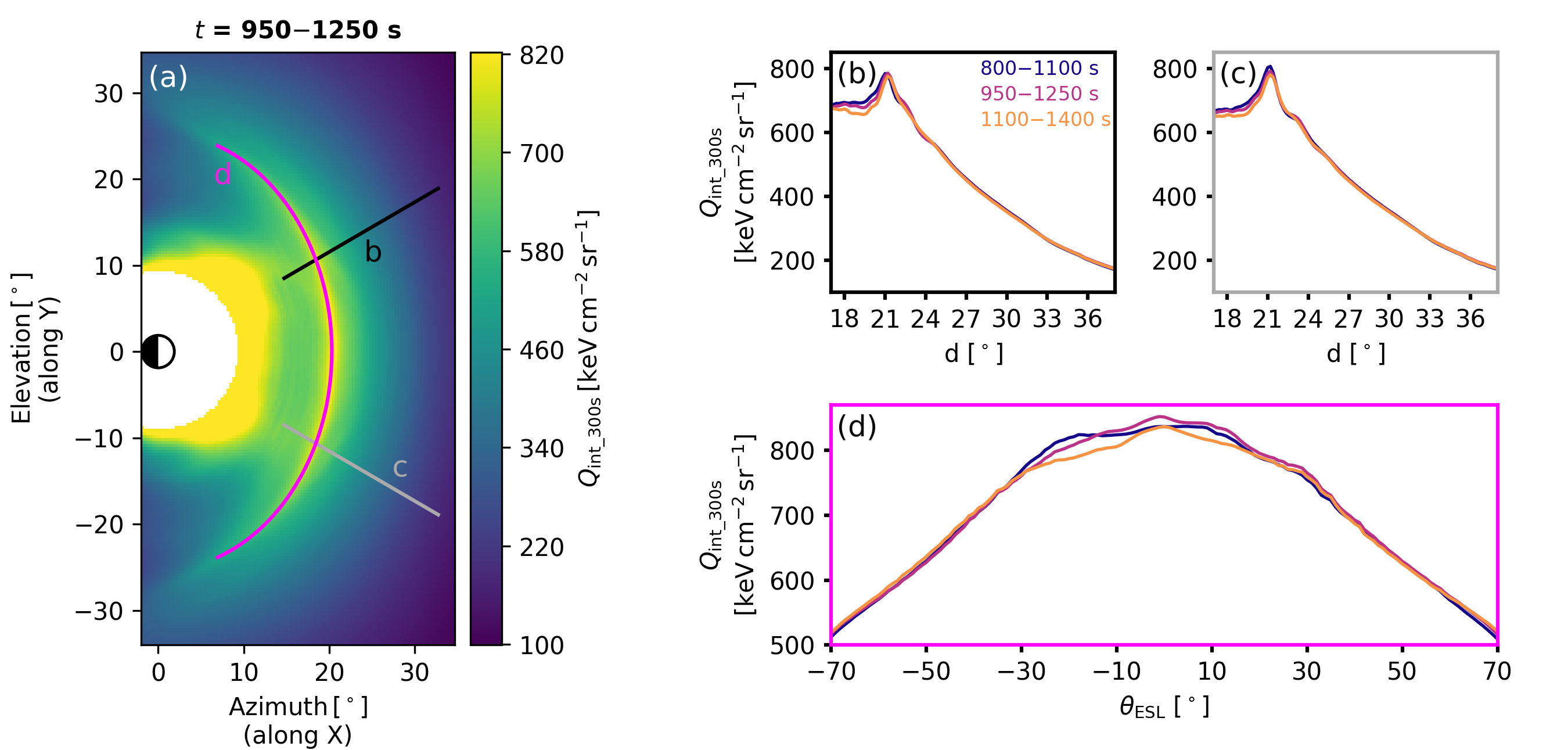}
   \caption{(a)~Time-integrated soft X-ray image from $t=950$ to $t=1250$~s obtained from a virtual spacecraft located at $(0, 0, 30\,R_\mathrm{E})$ (same as Fig.~\ref{fig:Qint300sXY}b), with several cuts along and across boundary region signatures. (b)~Soft X-ray emission along the black cut through the post-noon magnetopause and magnetosheath signatures. (c)~Same along the grey cut through the pre-noon magnetopause and magnetosheath signatures. (d)~Same along the magenta cut following the dayside magnetopause signature. In panels b--d, the three lines correspond to the three studied 300~s time intervals for image acquisition.}
   \label{fig:cutsXY}
\end{figure}

Figure~\ref{fig:cutsXY} shows the values of $Q_\mathrm{int\_300s}$ along a few selected cuts, in a same way as in Fig.~\ref{fig:cutsXZ}. Figure~\ref{fig:cutsXY}a reproduces Fig.~\ref{fig:cutsXZ}b and indicates where three cuts are considered to study the variations of  $Q_\mathrm{int\_300s}$ along the magnetopause X-ray signature as well as across the magnetosheath in the pre-noon and post-noon sectors. 

We can see from Figs.~\ref{fig:cutsXY}b--c that, along the cuts crossing the magnetosheath, the integrated soft X-ray emissivity peaks at \mbox{$d \approx 21^\circ$} in both the post-noon and pre-noon sectors, respectively, with a slight reduction in the peak value. Sunward from the peak (i.e. at larger $d$ values), fluctuations in $Q_\mathrm{int\_300s}$ can be seen, almost reaching a secondary peak at \mbox{$d \approx 23^\circ$} (more visible in the dawnside cut, Fig.~\ref{fig:cutsXY}c). This corresponds to the oblique stripes indicated with magenta arrows in Fig.~\ref{fig:Qint300sXY}.

Finally, looking at $Q_\mathrm{int\_300s}$ values along the magenta cut following the magnetopause signature (Fig.~\ref{fig:cutsXY}d), we note that the signal roughly follows linear slopes in the flanks \mbox{($|\theta_\mathrm{ESL}| > 30^\circ$)}, with little to no evolution with time. On the other hand, the section corresponding to the nose of the magnetopause exhibits fluctuations and variability amounting to about 5\% of the $Q_\mathrm{int\_300s}$ values (up to 40~keV\,cm$^{-2}$\,sr$^{-1}$ between the violet and the orange lines at $\theta_\mathrm{ESL} \approx -20^\circ$). In a given image (i.e., focusing on a single line in Fig.~\ref{fig:cutsXY}d), fluctuations can lead to the appearance of several local maxima in $Q_\mathrm{int\_300s}$, denoting the existence of structures near the magnetopause nose. These structures might be associated with the nature of the reconnection taking place at the dayside magnetopause, which can either be over an extended X-line or on the contrary more patchy \cite{Atz2022,Walsh2017}.

\section{Signatures of Transient Processes in Soft X-ray Images}
\label{sec:transient}

In this section, we will investigate to what extent transient processes taking place in the magnetosheath and at the dayside magnetopause can induce signatures in soft X-ray images as will be measured by SMILE and LEXI SXI. We will focus on two processes: flux transfer events forming by magnetic reconnection at the dayside magnetopause and mirror-mode waves developing in the magnetosheath.

\subsection{Flux Transfer Events at the Magnetopause}
\label{sec:FTE}

\begin{figure}
   \noindent\includegraphics[width=\textwidth]{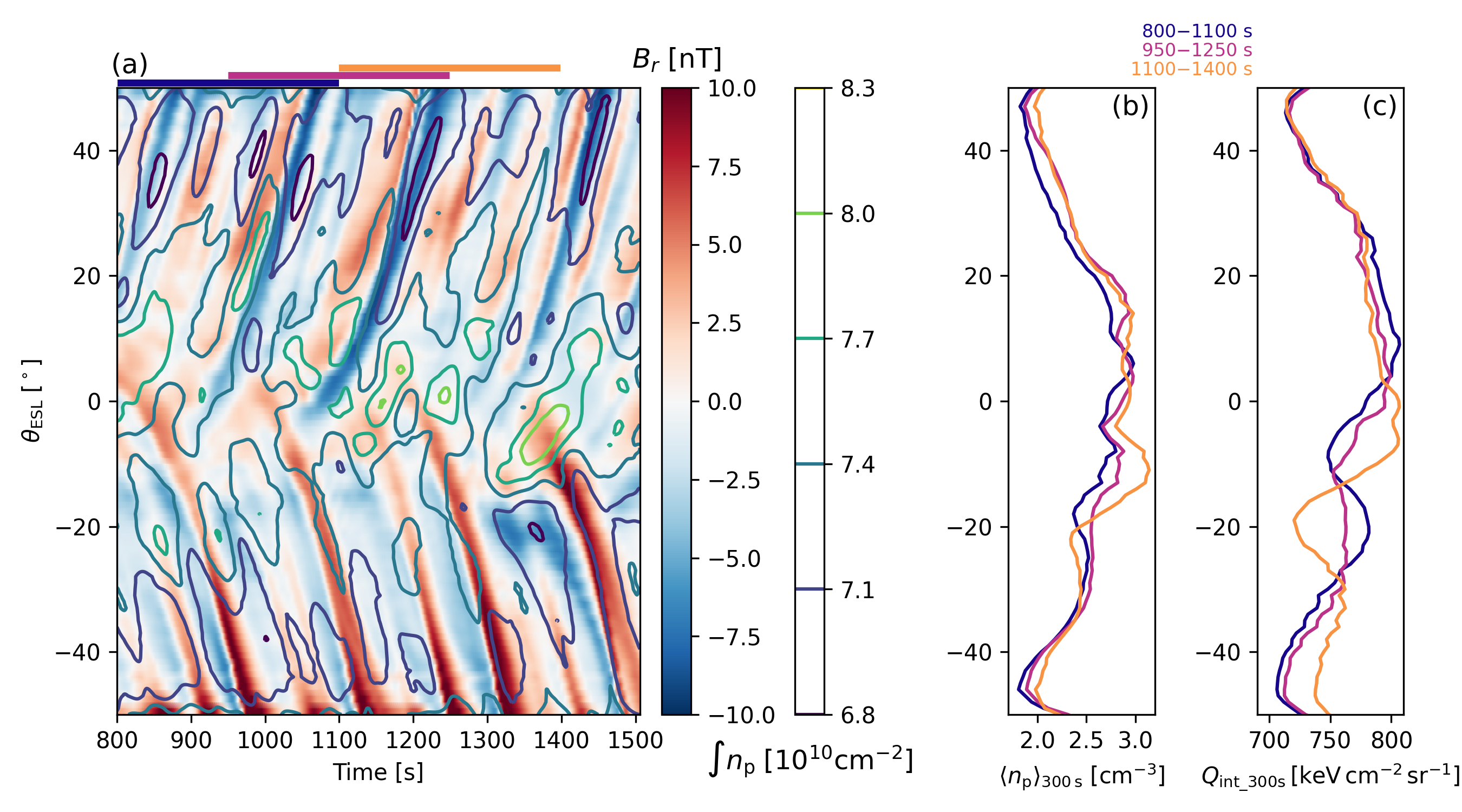}
   \caption{(a)~Background color: Time evolution of the radial (i.e., nearly normal to the magnetopause) component of the magnetic field along the intersection of the magenta cut following the magnetopause signature in Fig.~\ref{fig:cutsXZ}a with the $Y=0$ plane. FTEs generated near the subsolar point create a bipolar signature moving toward higher magnetopause angles as a function of time, with opposite polarity in opposite hemispheres. Isocontours: Time evolution of the line-of-sight-integrated proton number density along the magenta cut in Fig.~\ref{fig:cutsXZ}a. (b)~Proton number density along the intersection of the cut with the $Y=0$ plane, averaged over 300~s for the three studied time intervals (starting at $t = 800$~s, $t = 950$~s and $t = 1100$~s). (c)~Value of the 300~s-integrated soft X-ray emission along the cut in the corresponding three images shown in Fig.~\ref{fig:Qint300sXZ} (same data as Fig.~\ref{fig:cutsXZ}f). The colored horizontal lines above panel~(a) indicate the three studied time intervals.}
   \label{fig:FTEs}
\end{figure}

In Vlasiator 2D--3V runs driven by IMF with a southward $B_z$ component, flux transfer events regularly form at the magnetopause near the subsolar point \cite{Hoilijoki2017, Hoilijoki2019}, and they propagate poleward toward the cusps, with consequences in terms of energy transfer into the magnetosphere \cite{AlaLahti2022} and proton precipitation into the cusps \cite{Grandin2020}.

In this 3D--3V run, FTEs also form near the subsolar point and gradually follow the magnetopause toward the cusps, in both hemispheres. This can be seen in Supplementary Movie~S3, where FTEs appear as regions of increased proton density enclosed within a magnetic island. To investigate whether these FTEs can have a signature on soft X-ray images, we first find a simple way to detect and track them along the dayside magnetopause, and we then look at their time-integrated parameters over 300~s windows.

Figure~\ref{fig:FTEs}a shows (background color), as a function of time in the simulation, the magnetic field component along the radial direction (away from the Earth), $B_r$, along the magenta cut following the magnetopause signature shown in Fig.~\ref{fig:cutsXZ}a. More precisely, the values of $B_r$ are taken along the intersection of the lines of sight forming this cut with the $Y=0$ plane (noon--midnight meridional plane), and they approximately correspond to the magnetic field component normal to the magnetopause surface. One can see that, at a given value of $\theta_\mathrm{ESL}$ along the intersected cut, $B_r$ exhibits pseudo-oscillations with amplitudes up to $\pm$10~nT. These correspond to bipolar signatures as FTEs transit across the location. In the northern hemisphere, when the leading edge of a FTE approaches a given location at the dayside magnetopause, $B_r$ is increased (red values in the plot), and as the trailing edge of the FTE passes the location, $B_r$ shows a negative deflection (blue values). Due to the geometry, the situation is opposite in the southern hemisphere. Hence, we can see how FTEs tend to form within a few degrees from the subsolar point at the dayside magnetopause, and how they propagate toward one of the polar cusps in $\sim$150~s. It is noteworthy that the propagation time of the identified FTEs is shorter than the image acquisition time which is retained in this study. However, one may want to determine whether FTEs can still create a signature in soft X-ray images integrated over 300~s.

In Fig.~\ref{fig:FTEs}b, we show the time-averaged value of the proton number density, $\left< n_\mathrm{p} \right>_\mathrm{300s}$, along the intersection of the cut with the $Y=0$ plane (i.e., at the locations where $B_r$ is monitored). The time averaging is done over 300~s intervals corresponding to the integration time of the three images shown in Fig.~\ref{fig:Qint300sXZ}. It can be seen that (i)~the time-averaged density along the magnetopause exhibits some variations as a function of $\theta_\mathrm{ESL}$, and (ii)~these variations change from one 300~s interval to another. The largest differences occur at $\theta_\mathrm{ESL}$ values comprised between $0^\circ$ and $-20^\circ$ during the third time interval (1100--1400~s). During this time interval, three bipolar signatures can be identified in Fig.~\ref{fig:FTEs}a at those $\theta_\mathrm{ESL}$ values. We can see in Movie~S3 that these FTEs are indeed associated with higher-than-background proton number densities.

Figure~\ref{fig:FTEs}c reproduces the corresponding values of $Q_\mathrm{int\_300s}$ during the three time intervals along the cut in the soft X-ray images (same data as in Fig.~\ref{fig:cutsXZ}f). While the match between the spatial and temporal variations in $\left< n_\mathrm{p} \right>_\mathrm{300s}$ and $Q_\mathrm{int\_300s}$ is not perfect, the corresponding curves do nonetheless exhibit striking resemblance. This suggests that the presence of FTEs at the dayside magnetopause during the image acquisition time can affect the average proton density along corresponding lines of sight and induce a signature in soft X-ray emission values. The imperfect match can be expected given that $Q_\mathrm{int\_300s}$ values result from a line-of-sight integration, whereas the $\left< n_\mathrm{p} \right>_\mathrm{300s}$ values are taken locally in the $Y=0$ plane. We note, however, that the FTEs in this Vlasiator run can extend several Earth radii along the $Y$ direction (see Fig.~S1 in the Supplementary Material). Hence, in this imaging geometry, increased soft X-ray emissivity can take place over a significant portion of the line of sight intersecting an FTE.

To investigate this further, overlaid with the $B_r$ data in Fig.~\ref{fig:FTEs}a are isocontours of the proton number density integrated along the lines of sight forming the magenta cut in Fig.~\ref{fig:cutsXZ}a. As for the $Q_\mathrm{int\_300s}$ shown in Fig.~\ref{fig:FTEs}c, this line-of-sight-integrated density likely contains effects of variations occurring outside of the $Y=0$ plane; however, one can see a tendency for density enhancements to follow the pattern formed by the bipolar signatures associated with FTEs in the $Y=0$ plane. This suggests that the proton density enhancements occurring in the core of the simulated FTEs can produce soft X-ray signatures even in line-of-sight and time-integrated images. It is worth mentioning that, in observations, FTEs often exhibit a density decrease in their core \cite{Akhavan-Tafti2018}. Yet, some observed FTEs do show a proton density increase and might therefore produce signatures in soft X-ray images similar to those described here (see Sect.~\ref{sec:discussion})

\subsection{Mirror-Mode Waves in the Magnetosheath}
\label{sec:mirrormode}

In Movie~S3, one can identify wave-like structures in the magnetosheath consisting of patches of increased proton number density appearing in the central part of the magnetosheath (Earthward from the bow shock) and slowly drifting toward the magnetopause. In 2D--3V Vlasiator simulations, mirror-mode waves have been identified in the magnetosheath \cite{Hoilijoki2016,Dubart2020}. In a nearly 3D--3V setup wherein a noon--midnight meridional-plane slice with a thickness of $7 R_\mathrm{E}$ along the dawn--dusk dimension was simulated with Vlasiator, \citeA{PfauKempf2020} identified anticorrelated magnetic field and proton density fluctuations in the magnetosheath, attributed to mirror-mode waves. We will first check whether in this 3D--3V run those waves exhibit similar properties as mirror-mode waves, after which we will investigate to what extent they can create signatures in soft X-ray images.

\begin{figure}
   \noindent\includegraphics[width=.9\textwidth]{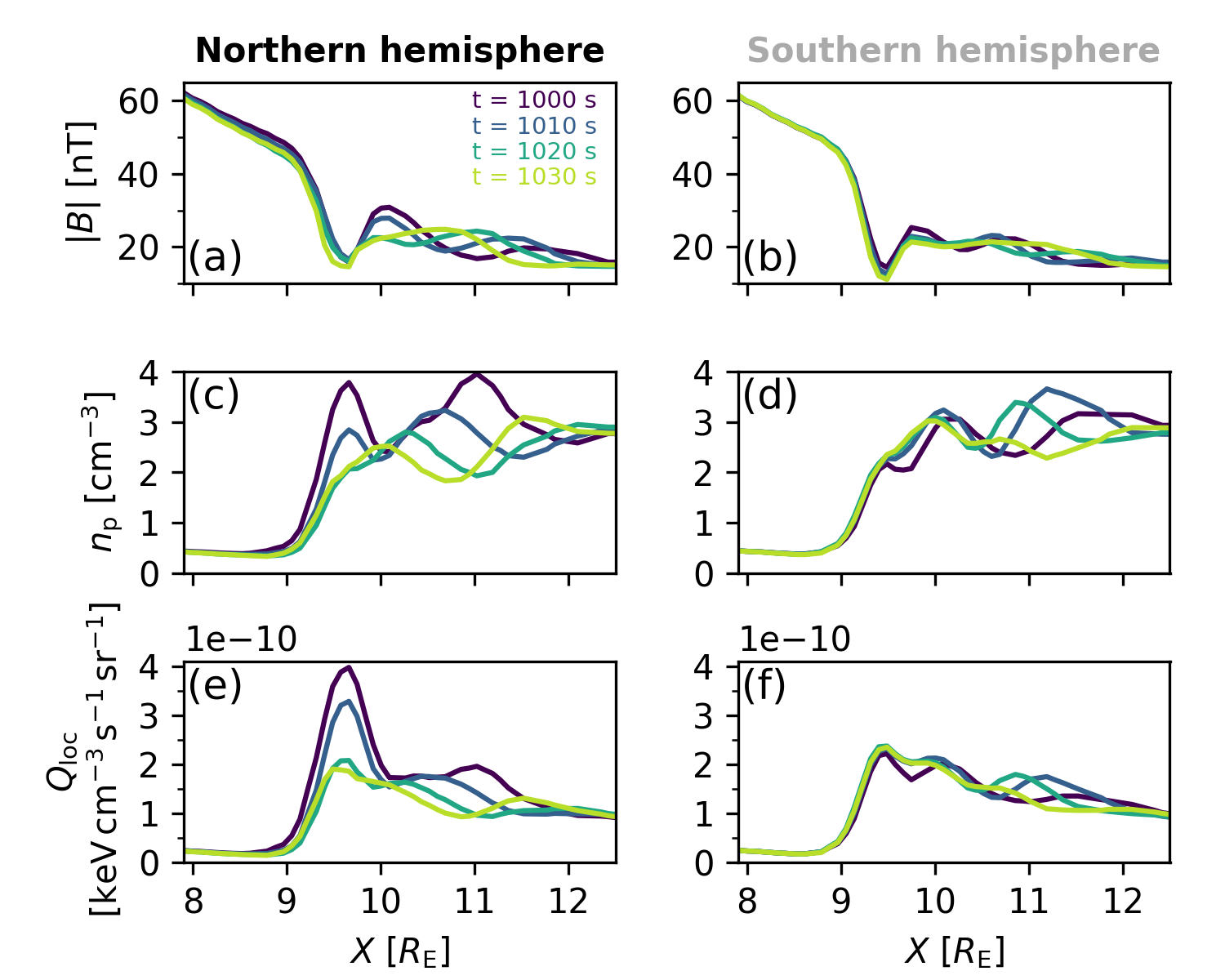}
   \caption{Investigation of the magnetosheath waves signatures along the black and grey cuts in Fig.~\ref{fig:cutsXZ}a. The shown variables are taken along the intersections of the cuts with the $Y=0$ plane at four time steps in the simulation (i.e., these are instantaneous values). (a--b)~Magnetic field magnitude. (c--d)~Proton number density. (e--f)~Local soft X-ray emissivity.}
   \label{fig:mirrorMode}
\end{figure}

Figure~\ref{fig:mirrorMode} presents the magnetic field magnitude (Fig.~\ref{fig:mirrorMode}a--b), proton number density (Fig.~\ref{fig:mirrorMode}c--d) and local soft X-ray emissivity (Fig.~\ref{fig:mirrorMode}e--f) along the intersections of the black and grey cuts in Fig.~\ref{fig:cutsXZ}a with the $Y=0$ plane. These data are shown at four time steps in the simulation: $t=1000$, 1010, 1020, and 1030~s. Here, the horizontal axis corresponds to the $X$ coordinate of the cut intersection points in the $Y=0$ plane to give a better intuition of spatial scales.

At $X<9\,R_\mathrm{E}$, the magnetic field magnitude decreases steadily and the proton density is on the order of 0.5~cm$^{-3}$, which corresponds to magnetosphere field and plasma. As $X$ increases, the magnetic field magnitude drops more drastically, whereas the proton density increases. This corresponds to the crossing of the magnetopause. Beyond the magnetopause, one can notice fluctuations in both $|B|$ and $n_\mathrm{p}$, with sizes on the order of 1--2\,$R_\mathrm{E}$ along the $X$ coordinate. These fluctuations are more prominent in the northern (black) cut than in the southern (grey) cut, and there is a fairly clear anticorrelation in the variations of both parameters. Besides, we can see that the structures associated with these fluctuations appear to drift Earthward by about 1\,$R_\mathrm{E}$ in 40~s, giving an estimated speed along the $X$ direction of $\sim$150~km\,s$^{-1}$. Comparing with Fig.~\ref{fig:snapshot_XZ}b, we can infer that the structures therefore approximately drift alongside the plasma. The anticorrelation between magnetic field and density variations and the fact that the waves are roughly static in the plasma frame strongly suggest that these waves are indeed mirror modes. The interhemispheric asymmetry mentioned above likely arises from the fact that small numerical effects (e.g., rounding errors) lead to mirror-mode structures not being exactly symmetrical in both hemispheres at a given time in the simulation.

\begin{figure}
   \noindent\includegraphics[width=.9\textwidth]{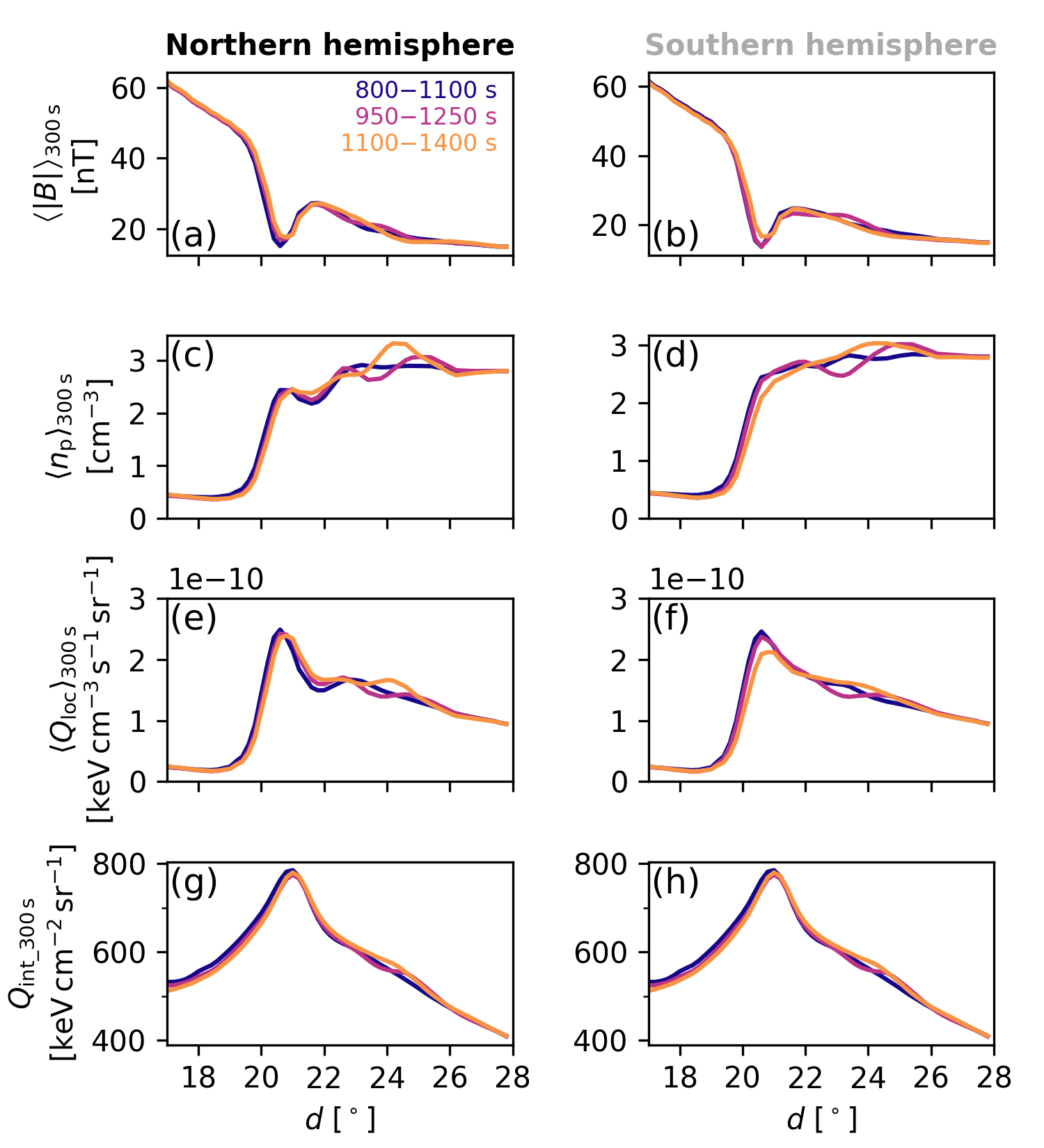}
   \caption{Time-averaged values of the magnetosheath parameters shown in Fig.~\ref{fig:mirrorMode} over 300~s intervals starting at $t=800$~s, $t=950$~s, and $t=1100$~s. (a--b)~Time-averaged magnetic field magnitude along the intersections of the black and grey cuts with the $Y=0$ plane, respectively. (c--d)~Same for the time-averaged proton number density. (e--f)~Same for the time-averaged local soft X-ray emissivity. (g--h)~Soft X-ray emission values along the black and grey cuts in the 300~s-integrated images shown in Fig.~\ref{fig:Qint300sXZ}. The last two panels show the same data as Fig.~\ref{fig:cutsXZ}b--c.}
   \label{fig:mirrorMode300s}
\end{figure}

Given the significant propagation of the mirror-mode waves over the span of 40~s seen in Fig.~\ref{fig:mirrorMode}, one cannot expect that individual wave fronts will be visible in an image integrated over 300~s. However, like for FTEs, one can investigate whether mirror-mode wave activity in the magnetosheath can nevertheless produce a signature in soft X-ray images. Figure~\ref{fig:mirrorMode300s} presents this analysis along the same cuts as Fig.~\ref{fig:mirrorMode}, but this time for time-averaged parameters along the cut, again using the same three 300~s time intervals as previously. In all panels of this figure, the horizontal axis corresponds to the angular distance $d$ to the origin of the azimuth--elevation frame used in soft X-ray images. 

In Figs.~\ref{fig:mirrorMode300s}a--b, we show the time-averaged magnetic field magnitude along the cuts during the three intervals. While the magnetosphere and magnetopause values do not differ significantly during the three intervals, one can see that the magnetosheath fluctuations, associated with the mirror-mode waves identified previously, differ slightly from each other. The differences appear more clearly in the time-averaged proton density panels (Figs.~\ref{fig:mirrorMode300s}c--d), especially at $d \approx 21$--$26^\circ$, and the resulting time-averaged local soft X-ray emission values ($\left< Q_\mathrm{loc} \right>_\mathrm{300s}$, Figs.~\ref{fig:mirrorMode300s}e--f) in the $Y=0$ plane are consistent with average density fluctuations. These asymmetries are however less prominent than those seen in Fig.~\ref{fig:mirrorMode}, because the 300~s averaging smooths them out. To assess the effect of these local time-averaged fluctuations in actual soft X-ray images, Figs.~\ref{fig:mirrorMode300s}g--h show the soft X-ray emissions integrated along the line of sight corresponding to the two cuts (same data as in Fig.~\ref{fig:cutsXZ}b--c, respectively). In these panels, we can see that the three lines corresponding to the three time intervals do not fully overlap within $d \approx 22$--$26^\circ$.

\section{Discussion}
\label{sec:discussion}

First of all, it is worth comparing the soft X-ray emissivity values obtained with the Vlasiator simulation to those reported in modeling studies using different models. \citeA{Sun2019} carried out a series of soft X-ray emission simulations using a MHD model, driven by various types of solar wind conditions. Our driving conditions correspond closest to their first case, with a solar wind density of 5~cm$^{-3}$, a solar wind speed of 400~km\,s$^{-1}$, and a purely southward IMF of 5~nT magnitude. They obtained a soft X-ray image from a virtual imaging satellite approximately positioned above the north pole (analogous to our Fig.~\ref{fig:snapshot_XY}d), in which the magnetopause signature reached a value of 13~keV\,cm$^{-2}$\,s$^{-1}$\,sr$^{-1}$. In contrast, with our Vlasiator run, we get a value close to 3~keV\,cm$^{-2}$\,s$^{-1}$\,sr$^{-1}$, which is almost a factor of 5 lower. This is consistent with our solar wind density being 5~times lower than theirs, leading to a magnetosheath density accordingly lower. Our values are also in fairly good agreement with those obtained by \citeA{Connor2021} using a simulation from the OpenGGCM global MHD model and providing soft X-ray images seen from the dawnside (analogous to our Fig.~\ref{fig:snapshot_XZ}d).

While Vlasiator is too computationally expensive to be run over a long time interval, its unique advantage is that the hybrid-kinetic description of the space plasma enables the formation of transient features which cannot emerge in MHD simulations. We can therefore simulate the possible effect of such phenomena in soft X-ray images of the near-Earth space as will be obtained by SMILE and LEXI. Considering first the case of FTEs, if we focus on the largest peak-to-trough variation of $Q_\mathrm{int\_300s}$ in Fig.~\ref{fig:FTEs}c, corresponding to the 1100--1400~s time interval and $\theta_\mathrm{ESL}$ angles between $-20^\circ$ and $0^\circ$, we obtain \mbox{$\Delta Q_\mathrm{int\_300s} \approx 90$~keV\,cm$^{-2}$\,sr$^{-1}$}, amounting to about 12\% of the background value along the dayside magnetopause. The corresponding spatial variations in proton density in the $Y=0$ plane amount to 0.81~cm$^{-3}$, on the order of 30\% of the background value. This can provide quantitative elements to assess whether a soft X-ray imager might be able to detect signatures associated with FTEs, depending on the specifications of the soft X-ray sensor.

FTEs have been found to form at the dayside magnetopause quasi-periodically appearing approximately once every 8~minutes \cite{Rijnbeek1984} preferably forming during southward IMF conditions \cite<e.g.,>{Berchem1984}. A statistical study of 55~FTEs using MMS observations found an average size of a subsolar FTE to be \mbox{$1700\pm400$~km} \cite{Akhavan-Tafti2018}, which is 3 to 7~times smaller than the FTEs observed at the higher latitudes \cite{Akhavan-Tafti2018, Fermo2011, Wang2005}. However, the scale sizes of the observed FTEs vary from ion-scale flux ropes \cite<e.g.,>{Eastwood2016, Dong2017} to FTEs with a diameter up to 1--2\,$R_\mathrm{E}$ \cite{Rijnbeek1984,Fear2007,Hasegawa2006}, while their axial length can be much longer \cite<e.g.,>{Fear2008}. 

A typical FTE has an internal structure with twisted field lines and axial magnetic field leading to an increase in total magnetic field magnitude and a decrease in the density in the core of the FTE \cite<e.g.,>{Akhavan-Tafti2018, Zhang2010}. However, ``crater'' FTEs with a decrease in total magnetic field with a higher density in the middle have been observed \cite{Zhang2010}. A statistical study of 18~typical and 14~crater FTEs comparing FTE densities ($n_\mathrm{FTE}$) and magnetic field strength ($B_\mathrm{FTE}$) to the magnetosheath density ($n_\mathrm{MS}$) and magnetic field ($B_\mathrm{MS}$) showed that 75\% of typical FTEs have the density ratio \mbox{$n_\mathrm{FTE}/n_\mathrm{MS}<0.5$} and 80\% of the FTEs have \mbox{$B_\mathrm{FTE}>B_\mathrm{MS}$}, while 90\% of the crater FTEs had \mbox{$n_\mathrm{FTE}/n_\mathrm{MS}>0.5$} (and approximately 50\% \mbox{$n_\mathrm{FTE}>n_\mathrm{MS}$}) and approximately 95\% had smaller magnetic field magnitude than the surrounding magnetosheath \cite{Zhang2010}. Recent particle-in-cell (PIC) simulations have revealed the important role of a guide field in FTEs. A local PIC study showed that, if the guide field is strong, FTEs exhibit low density in their core, whereas a weak guide field results in enhanced proton density within the FTE \cite{Lu2020}. These results have been confirmed in a global PIC study, wherein purely southward IMF ($B_y = 0$, no guide field) leads to density increase in FTEs whereas driving conditions with a nonzero IMF $B_y$ (i.e. with a guide field) lead to FTEs exhibiting a density decrease in their core \cite{Guo2021}. In this Vlasiator simulation, the IMF is purely southward, and the identified FTEs have a higher-than-background proton density in their core. This result is consistent with those from \citeA{Guo2021}. We can therefore speculate that soft X-ray images might be able to capture signatures of FTEs when their guide field is weak ($B_y \approx 0$), even more so if their peak-to-trough density variation is similar to or greater than that of FTEs identified in this simulation. However, we can anticipate that FTEs forming in presence of a strong guide field ($B_y \neq 0$) would not produce bright signatures in soft X-ray images.

It is clear from the presented results that FTEs forming in this Vlasiator run move across a 300~s integrated soft X-ray image too fast for them to be individually visible in the image. Rather, the local soft X-ray emission enhancements in some parts of the magnetopause signature are likely due to the cumulative effect of several FTEs, especially if these tend to remain longer at a given latitude. \citeA{Hoilijoki2019} showed that the speed at which FTEs propagate toward higher latitudes along the magnetopause depends on the driving conditions. We can therefore speculate that, in some cases where the FTE motion is slower than in this simulation, their signatures in soft X-ray images might be more prominent than in the results obtained here.

Regarding mirror-mode waves in the magnetosheath, we can estimate the effect of the largest density enhancement visible in Fig.~\ref{fig:mirrorMode300s}c. Considering the difference between the $\left< n_\mathrm{p} \right>_\mathrm{300s}$ values during the first and third time intervals at $d=24^\circ$, we find that it consists of a local density enhancement by 0.4~cm$^{-3}$ ($\sim$14\%) in the $Y=0$ plane associated with mirror-mode waves structures. The corresponding difference in time-integrated soft X-ray emissions amounts to about 20~keV\,cm$^{-2}$\,sr$^{-1}$ ($\sim$4\%).

Mirror-mode structures are large-scale compressional waves that are non-propagating in the plasma frame. They are often observed in the Earth’s magnetosheath especially behind the quasi-perpendicular bow shock. They may appear as quasi-sinusoidal oscillations in the magnetic field but mirror-mode structures are often observed as peaks in the mirror-unstable plasma and dips in mirror-stable conditions near the magnetopause \cite{Soucek2008}. The depth of the mirror structures has been found to increase with the decreasing distance to the magnetopause \cite{Soucek2008}. The analysis of 2~months of Cluster data by \citeA{Soucek2008} showed that the average period of the mirror-mode waves was approximately 12~s (with 98\% of the observed structures being between 4 to 24~s). A statistical study by \citeA{Lucek2001} showed that the scale sizes of the mirror structures can vary from less than 600~km along the local magnetopause normal, 750--1000~km along the maximum variance direction to 1500--3000~km along the flow direction. The amplitude of the magnetic field variations associated with mirror-mode waves is typically on the order of several 10\% of the background field, and the associated proton density variations can lead to up to multi-fold local enhancements \cite<e.g.,>{Chandler2021}.

The differences in $Q_\mathrm{int\_300s}$ values obtained above are estimates under the driving conditions and the setup used in the Vlasiator run on which this study is based. These choices and constraints have several consequences on the properties of the mirror-mode waves. First of all, in our simulation, the spatial resolution in the magnetosheath (2000~km) likely leads to mirror-mode structures being larger than in reality. Therefore, one may speculate that SMILE and LEXI observations of mirror-mode structures would likely be smaller and might hence be blurred out more easily than in this simulation.

Besides, \citeA{Dubart2020} showed that, when the spatial resolution of a Vlasiator simulation is too coarse, the unstable proton distributions in the magnetosheath entirely pump their free energy into the growth of mirror-mode structures, instead of transferring part of it to electromagnetic ion cyclotron waves, which require a finer spatial resolution to emerge. It is worth noting, however, that the solar wind driving conditions used in the run are not the most typical ones. Indeed, the solar wind density is notably lower than average, whereas the solar wind velocity and temperatures are higher than average. This suggests that, under solar wind conditions closer to average values than in this Vlasiator run, such as those from \citeA{Sun2019} discussed above, larger signal than what we obtained can be expected. This, combined with the fact that density fluctuations associated with mirror-mode waves can be greater than in our simulation (factor of 3 in the event analyzed by \citeA{Chandler2021}), suggests that the constraints on FTEs and mirror-mode waves for them to produce signatures in the soft X-ray observations of LEXI and SMILE might actually be less strict than suggested by our results. Additional simulations with a finer spatial resolution and different solar wind driving are needed to test this hypothesis.

Another point worth discussing is that the Vlasiator run duration analyzed for this study consists of a 700~s time interval only. In reality, both LEXI and SMILE SXI instruments will observe processes over significantly longer time scales, which could open more possibilities for the data analysis. For instance, calculating the discrepancy between two time-integrated images separated by a few minutes could enhance contrast in regions where the plasma conditions have changed during that time frame. While this would likely still not enable the detection and monitoring of individual structures given the required image acquisition time, this could provide additional approaches to indirectly identify such processes in near-Earth space, as well as context for in-situ observations.

Finally, one should point out the fact that, in this Vlasiator run, the driving conditions are steady. As a consequence, one may expect that the large-scale features such as the boundary locations (bow shock, magnetopause, polar cusps) do not change much during the studied time interval. Nevertheless, Figs.~\ref{fig:cutsXZ}d--e indicate that the cusp signatures slightly move toward higher latitudes with time, possibly in response to processes occurring on the nightside in this simulation (not shown). This suggests that the analysis of successive soft X-ray images might enable the tracking of boundary motions. To confirm this, it would prove interesting to monitor the motion of these boundaries in a Vlasiator run in which solar wind conditions are varying. This is a future avenue for an upcoming study which could be made possible once a Vlasiator run with time-varying driving conditions is available.

\section{Conclusions}
\label{sec:conclusion}

In this paper, we presented the first estimates of SWCX soft X-ray emission imaging of the dayside near-Earth space based on a hybrid-Vlasov simulation. We used $\sim$700~s of a Vlasiator run driven by constant solar wind of low density and high velocity and with purely southward IMF with $B_z = -5$~nT. From the Vlasiator simulation outputs, we calculated at every second the local soft X-ray emissivity based on the plasma parameters and produced line-of-sight-integrated instantaneous soft X-ray images as would be obtained from a virtual imaging spacecraft placed at a distance of $30\,R_\mathrm{E}$ from the Earth's center, providing the polar and side views of the Earth's magnetosphere similar to the SMILE and LEXI views, respectively. We then integrated those instantaneous images over 300~s intervals to reproduce anticipated observational requirements for the SMILE and LEXI SXI instruments. The main results of this study are as follows:
\begin{enumerate}
    \item The most prominent features in soft X-ray images obtained from above the pole or on the dawnside consist of the magnetopause and polar cusps (the latter only when observing from the dawnside). The soft X-ray emissivity values obtained with Vlasiator are consistent with earlier MHD results, taking into account differences in the solar wind driving conditions.
    \item Despite the 300~s integration time, the obtained soft X-ray images exhibit smaller-scale features such as the brightening of localized areas at the dayside magnetopause and wave-like patterns in the magnetosheath signature. These features are created by transient phenomena, namely FTEs and mirror-mode waves, respectively.
    \item Based on the simulated soft X-ray images, one can anticipate that FTE signatures could amount to 12\% of the background signal if FTEs occurring during the 300~s interval cumulatively lead to local proton density enhancements by about 30\%. This result holds for FTEs whose guide field is weak (when IMF $B_y \approx 0$), which exhibit enhanced proton density in their core \cite{Lu2020, Guo2021}. One can expect that FTEs with a strong guide field, which contain low-density plasma, would not produce bright signatures in soft X-ray images.
    \item Correspondingly, soft X-ray images of SMILE and LEXI might reveal the presence of mirror-mode waves, as in the simulation the associated density structures in the magnetosheath cumulatively led to an enhancement by 14\% in the proton density, locally, and produced an increase in soft X-ray emission signal by 4\%. Depending on the scale size, propagation speed and amplitude of the variations in the proton density, mirror-mode structures may lead to more pronounced or on the contrary more blurred out signatures in soft X-ray images.
    \item The above values are likely conservative estimates, given that the solar wind conditions in this Vlasiator run are representative of solar wind high-speed streams rather than the more common average solar wind driving which typically has larger proton number density.
\end{enumerate}

The results of this study contribute to the SMILE Modeling Working Group objectives by comparing simulation results using a hybrid-Vlasov code with other works based on MHD. Moreover, the hybrid-kinetic description of the space plasma provides a unique opportunity to anticipate under what conditions transient processes in the dayside near-Earth space might be observed by soft X-ray imagers on SMILE and LEXI. Future plans could include the monitoring of the motion of the large-scale space plasma boundaries (magnetopause, polar cusps, bow shock) in a different run with time-varying upstream conditions (for instance a dynamic pressure increase or a rotation of the IMF).


\acknowledgments
We acknowledge the European Research Council for starting grant 200141-QuESpace, with which the Vlasiator model was developed, and consolidator grant 682068-PRESTISSIMO awarded for further development of Vlasiator and its use in scientific investigations. We gratefully acknowledge Academy of Finland grant numbers 338629-AERGELC'H, 339756-KIMCHI, 336805-FORESAIL, and 335554-ICT-SUNVAC. The Academy of Finland also supported this work through the PROFI4 grant (grant number 3189131). Hyunju K. Connor gratefully acknowledges support from the NASA grants, 80NSSC20K1670 and 80MSFC20C0019, and the NASA GSFC FY23 IRAD and HIF funds. The CSC -- IT Center for Science and the PRACE Tier-0 supercomputer infrastructure in HLRS Stuttgart (grant number 2019204998) are acknowledged as they made these results possible. The authors wish to thank the Finnish Grid and Cloud Infrastructure (FGCI) for supporting this project with computational and data storage resources.

The authors declare that they have no conflict of interest.

The Vlasiator simulation data used in the study amount to 18~TB of disk space; access to the raw simulation data can be granted by following the Vlasiator data access policy (see \url{https://www2.helsinki.fi/en/researchgroups/vlasiator/rules-of-the-road}).
The Vlasiator code is preserved at \url{https://zenodo.org/record/3640593} \cite{Vlasiator_code}, available in open access. It is developed openly at \url{https://github.com/fmihpc/vlasiator}.


%
\bibliography{smile_vlasiator_bibliography}
%




%
%
%
%
%

\end{document}